\documentclass[reprint,amsmath,amssymb,nofootinbib,aps,pra, preprintnumbers]{revtex4-2}
\pdfoutput=1

\usepackage{graphicx}
\usepackage{color}
\usepackage[dvipsnames]{xcolor}
\usepackage{xspace}
\usepackage{multirow}
\usepackage[normalem]{ulem}
\usepackage{makecell}
\usepackage{appendix}
\usepackage[breaklinks=true,colorlinks=true,linkcolor=blue,urlcolor=blue,citecolor=blue]{hyperref}

\def\apj{ApJ}

\begin{document}

\title{Relaxing CMB bounds on Primordial Black Holes: the role of ionization fronts}

\author{Ga\'etan Facchinetti}
\author{Matteo Lucca}
\author{Sebastien Clesse}
\affiliation{Service de Physique Th\'{e}orique, Universit\'{e} Libre de Bruxelles, C.P. 225, B-1050 Brussels, Belgium}

\begin{abstract}
	The accretion of matter onto primordial black holes (PBHs) during the dark ages and the subsequent energy injection in the medium should have left imprints on the cosmic microwave background (CMB) anisotropies. Recent works have claimed stringent CMB limits on the PBH abundance, hardly compatible with a PBH interpretation of the gravitational-wave observations of binary BH mergers. By using a more realistic accretion model based on hydrodynamical simulations and conservative assumptions for the emission efficiency, we show that CMB limits on the PBH abundance are up to two orders of magnitude less stringent than previously estimated between $10$ and $10^4$~M$_\odot$. This reopens the possibility that PBHs might explain at the same time (at least a fraction of) the dark matter, some of the LIGO-Virgo-KAGRA binary BH mergers and the existence of super-massive BHs. More generally, we emphasize that PBH accretion can be a rather complex physical process with velocity dependences that are hard to assess, which introduces large uncertainties in accretion-based limits on the PBH abundance.
\end{abstract}

\maketitle
\preprint{ULB-TH/22-27}

\section{Introduction}

Should they exist, primordial black holes (PBHs) would be an invaluable probe of new physics. For instance, an unequivocal detection of even one single PBH would have fundamental consequences for our understanding of the primordial universe, shedding light for instance on the shape of the primordial power spectrum at very small scales -- inaccessible to probes such as the cosmic microwave background (CMB). At the same time, PBHs could also make up for a sizable fraction of the dark matter (DM) content of the universe (if not its totality) \cite{Carr2016Primordial, Carr2020Primordial}, further motivating the search for their existence.

In light of these considerations, the detection of almost 100 binary black hole (BH) mergers by the LIGO-Virgo-KAGRA (LVK) collaboration \cite{LIGOScientific:2021djp,LIGOScientific:2021usb,LIGOScientific:2020ibl,LIGOScientific:2018mvr}, some of which with intriguing properties such as low spins, BHs in the pair instability mass gap~\cite{LIGOScientific:2020iuh} and with low mass ratios~\cite{LIGOScientific:2020zkf}, has led to various analyses investigating whether the presence of PBHs is compatible with these gravitational-wave (GW) observations~\cite{Bird:2016dcv,Clesse:2016vqa,Sasaki:2016jop}. As it turns out, evidence in favour of the PBH hypothesis has been found if their density is between two and four orders of magnitude lower than that of the DM for PBH masses of the order of $10-100$ M$_\odot$ \cite{Clesse:2017bsw,Raidal:2017mfl,Raidal:2018bbj,Carr:2019kxo,Gow:2019pok,Hall:2020daa,Hutsi:2020sol,Clesse:2020ghq,DeLuca:2020qqa, DeLuca:2020sae,Jedamzik:2020omx,Jedamzik:2020ypm,Wong:2020yig,Escriva:2022bwe,Franciolini:2022tfm}. Interestingly, however, a large fraction of this region of the parameter space is currently excluded by CMB constraints, depending on the accretion model assumed \cite{2007ApJ...662...53R, 2008ApJ...680..829R, AliHaimoud2017Cosmic, Poulin2017CMB, Aloni:2016kuh, Serpico2020Cosmic, Juan:2022mir, Piga:2022ysp} (see e.g., Fig. 8 of \cite{Piga:2022ysp} for a graphical representation of the uncertainties involved). 

If present, PBHs would inevitably accrete matter in their surroundings, which would then heat up adiabatically and emit high energy radiation into the universe. This injection of energy can in turn impact a number of observables including the CMB anisotropy power spectra, which then allows to constrain the properties of these compact objects. Nevertheless, the predicted amount of emission, and hence the strength of the constraints, strongly depends on the assumptions made to describe the system. Most notably these involve the geometry of the accretion \cite{AliHaimoud2017Cosmic, Poulin2017CMB}, the temperature profile close to the BH \cite{AliHaimoud2017Cosmic} and the relative velocity between PBHs and the surrounding environment. The presence of outflows \cite{Piga:2022ysp} or DM halos \cite{2007ApJ...665.1277M, Serpico2020Cosmic} might further significantly affect the amount of emitted radiation (see e.g., Sec. 2.4 of \cite{Piga:2022ysp} for a more complete list of effects and references).

Therefore, in order to determine the compatibility of the evidence for PBHs derived from the LVK data with the constraints imposed by CMB anisotropies it is fundamental to develop a realistic and inclusive treatment of the relevant accretion physics. While, so far, mostly idealized models have been brought forward in the literature, several steps towards a more complete picture have been taken. 

For instance, in \cite{Park:2012cr} (PR13) the authors considered more closely the dependence of the interaction between PBHs and the surrounding medium as a function of the PBH proper velocity, and found that for realistic values of the latter one witnesses the formation of a dense cometary-shaped ionization front (I-front) ahead of the PBH. The presence of this shell reduces the density of the incoming flow (which gets in part tangentially deflected) so that the accretion rate is suppressed with respect to the case without I-fronts. This conclusion has been validated on the basis of various dedicated numerical simulations~\cite{Park:2012cr, Sugimura:2020rdw} and implies that accounting for this radiative feedback effect would suppress the amount of energy injected into the universe by the PBH accretion of matter. However, the impact of this effect on the related cosmological constraints on the PBH abundance has not been investigated so far.

In this work we build on the results presented in \cite{Park:2012cr, Sugimura:2020rdw} and analyse their cosmological implications focusing in particular on the CMB anisotropies. We derive state-of-the-art CMB bounds based on the PR13 model and compare them to the landscape of previously claimed limits from \cite{AliHaimoud2017Cosmic, Poulin2017CMB}. As a result, we find that the parameter space opens up by two or more orders of magnitude for PBH masses between $10$ and $10^4$ M$_\odot$. This clearly has very relevant implications for the LVK preferred window, which would be allowed by a quite large margin should the PR13 model reflect reality. Our results also have a broader impact when considering extended PBH mass functions that could explain at the same time a large DM fraction and the seeds of supermassive BHs~\cite{Carr:2019kxo}, a conjecture claimed to be excluded due to CMB limits~\cite{Juan:2022mir}.

This paper is organized as follows. In Sec.~\ref{sec: PR} we introduce the PR13 model focusing on its impact on the accretion rate, the emission efficiency and the CMB anisotropy power spectra. In Sec.~\ref{sec: num} we present the numerical pipeline employed to compute the relevant cosmological quantities and impose the CMB constraints, which we then discuss in Sec.~\ref{sec: res}. We conclude in Sec.~\ref{sec: conc} with a summary of our findings. 

\section{The Park-Ricotti model}\label{sec: PR}

We begin our analysis by summarizing the PR13 model introduced and developed in \cite{Park:2012cr, Sugimura:2020rdw}. In particular, we detail the predicted accretion rate in Sec. \ref{sec: acc}, the consequent luminosity in Sec. \ref{sec: eff} and the final impact on the CMB anisotropy power spectra in Sec. \ref{sec: imp}. For convenience, we use natural units ($G = c = 1$) in the rest of the paper.

\subsection{Accretion rate}\label{sec: acc}
The accretion rate of a static PBH of mass $M_{\rm PBH}$ in the cosmological medium is given in full generality by the Bondi accretion rate, which is defined as
\begin{align}
    \dot{M}_{\rm PBH} = 4\pi \rho_\infty\,\lambda \frac{M_{\rm PBH}^2}{c_{\infty}^3}\,,
\end{align}
where $\rho_\infty$ and $c_\infty$ are the baryon density and sound speed far away from the PBH, respectively, and $\lambda$ accounts for all deviations from this idealized scenario. Yet, PBHs are expected to have the same velocity as the DM-baryon linear relative velocity at large scales $v_L$ \cite{AliHaimoud2017Cosmic, Inman2019Early},
\begin{align}\label{eq: v_L}
    \langle v_L^2 \rangle^{1/2} = \text{min}[1, (1+z)/10^3] \times 30\,\text{km/s}\,,
\end{align}
so that the PBH proper velocity $v_{\infty}=\langle v_L^2 \rangle^{1/2}$ (to follow the same notation as PR13) needs to be taken into account. For simplicity, one could follow the Bondi-Hoyle-Lyttleton (BHL) model which predicts
\begin{align}
    \lambda = \frac{c_\infty^3}{(c_\infty^2+v_\infty^2)^{3/2}}\,.
\end{align} 
Nevertheless, as mentioned above, PR13 suggests that a more realistic treatment is required. 

The PR13 model defines three possible regimes depending on the velocity of the PBH, referred to as ``low'', ``intermediate'' and ``high'' velocity. In all cases, an I-front is assumed to form ahead of the PBH, distinguishing the neutral region ahead of the front (i.e., at infinity) from the region behind it (i.e., embedding the PBH) where the medium has been ionized. They are referred to by the state of the Hydrogen they contain, respectively 'I' (for H$_{\rm I}$) and 'II' (for H$_{\rm II}$). When the PBH is slow, the surrounding matter has the time to form a dense shell ahead of the I-front (D-type front, for ``dense''). On the contrary, when the PBH is moving with too high velocities, the shell does not have the time to form (R-type front, for ``rarefied''). The presence (or absence) of this shell then affects the velocity and the density of the matter entering the accretion region, introducing therefore a dependence of the accretion rate on the I-front type, which in turn depends on the velocity of the PBH.

Concretely, assuming the width of the shell to be negligible, one can impose conservation of energy between the two regions, from which follows that
\begin{align}
    \frac{\rho_{\rm II}}{\rho_{\rm I}}=\frac{v_{\rm I}}{v_{\rm II}}\equiv\Delta^{\rm D/R}\,,
\end{align}
where the jump conditions $\Delta^{\rm D/R}$ (referring to the R or D regime depending on the value of $v_\infty$) are defined in Eq. (1) of \cite{Sugimura:2020rdw} and respect the condition ${\Delta^{\rm D}>\Delta^{\rm R}\geq1}$, as expected. Furthermore, from this equation one can derive two critical velocities from the sound speeds: $v_{\rm D} \equiv c_{\rm I}^2/(2c_{\rm II})$, below which a D-type front forms, and $v_{\rm R} \equiv 2c_{\rm II}$, above which an R-type front forms. Because of the higher temperatures in the ionized region, the sound speeds satisfy $c_{\rm II} > c_{\rm I}$ and thus $v_{\rm D } < v_{\rm R}$. In the high and low velocity scenarios (i.e., when $v_\infty>v_{\rm R}$ and $v_\infty<v_{\rm D}$, respectively) one then has that 
\begin{align}
    \lambda = \frac{\rho_{\rm II}}{\rho_{\infty}}\left(\frac{c_{\infty}}{c_{\rm II}}\right)^3 = \frac{ \Delta^{\rm D/R} c_\infty^3}{\left[c_{\rm II}^2+(v_\infty/\Delta^{\rm D/R})^2\right]^{3/2}}\,,
\end{align}
since for the accretion one has to consider the environment close to the PBH, i.e., region II. For very high velocities, i.e., $v_\infty \gg c_{\rm II} > c_{\rm I}$, $\Delta^{\rm R}\to1$ and therefore $\lambda$ tends to the BHL case. On the other hand, in the intermediate velocity scenario (i.e., when $v_{\rm R}>v_\infty>v_{\rm D}$), one has that $\rho_{\rm II}=\rho_\infty(c_\infty^2+v_\infty^2)/(2c_{\rm II}^2)$, leading to
\begin{align}
    \lambda = \frac{\left(c_\infty^2+v_\infty^2 \right) c_\infty^{3}}{4\sqrt{2}\,c_{\rm II}^5}\,.
\end{align}
\begin{figure*}[t]
    \centering
    \includegraphics[width=0.92\columnwidth]{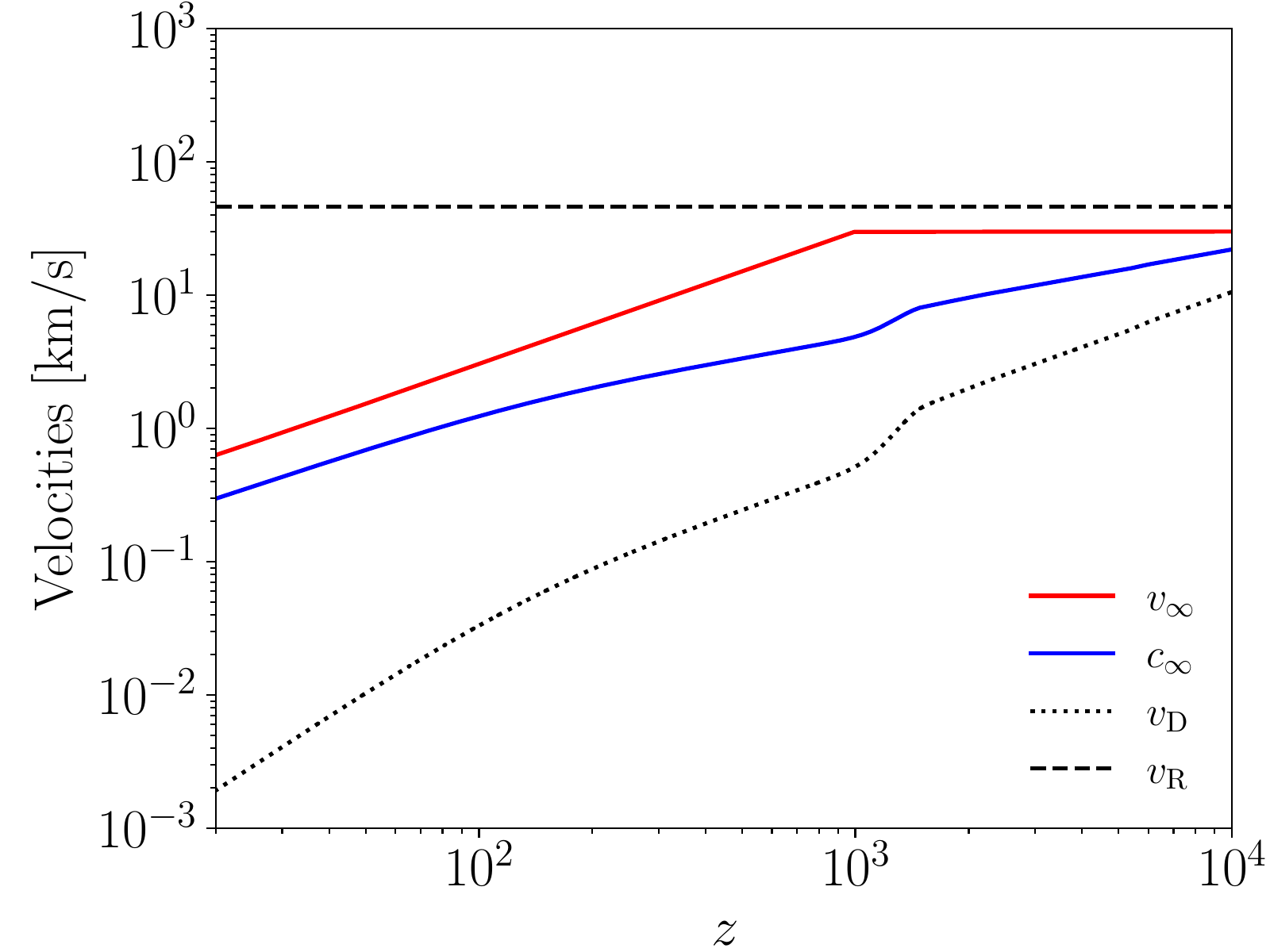}
    \includegraphics[width=0.92\columnwidth]{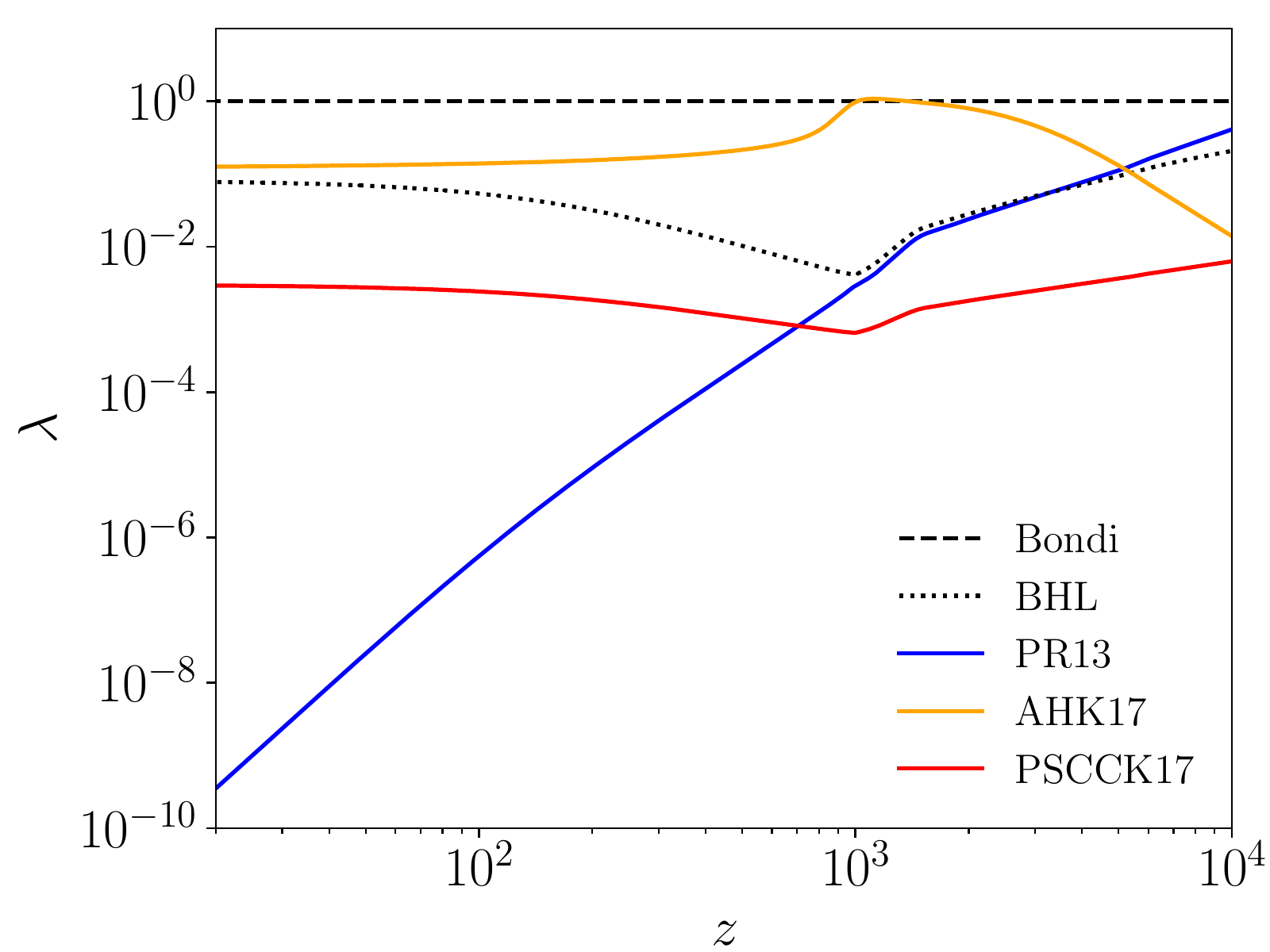}
    \caption{\textit{Left panel}: Characteristic velocities of the PR13 model. \textit{Right panel}: Corresponding redshift evolution of the accretion rate $\lambda$ (blue curve), together with the predictions of the same quantity in the Bondi (dashed black) and BHL (dotted black) limits as well as in the context of the spherical (orange) and disk (red) accretion scenarios presented in AHK17 \cite{AliHaimoud2017Cosmic} and PSCCK17~\cite{Poulin2017CMB}, respectively. In both plots we assume $M_{\rm PBH}=10^2$ M$_\odot$ and $c_{\rm II}=23$~km/s.}
    \label{fig: vel_lambda}
\end{figure*}

Various representations of the dependence of $\dot{M}_{\rm PBH}$ and $\lambda$ on $v_\infty$ are provided in e.g., Fig. 1 of \cite{Scarcella:2020ssk}, where different curves for different choices of energy density, $c_{\rm II}$ and PBH mass are shown, as well as in \cite{Park:2012cr, Sugimura:2020rdw} (note, however, that in the two references different values of $c_\infty$ are used). In reality, however, both the sound speed of the baryons and the PBH proper velocity are fixed by the cosmological model (with the latter defined as in Eq.~\eqref{eq: v_L}). Furthermore, the dependence of $\lambda$ on $M_{\rm PBH}$ is very weak as it enters only via $c_\infty$, which in turn depends on the baryon temperature that increases the more energy is injected in the system (the same also applies to the fractional PBH abundance $f_{\rm PBH}$, see below, which we henceforth set to unity for simplicity). We report therefore the chosen value of $M_{\rm PBH}$ to be exact, but note that the discussion carried out in this section does not significantly depend on it.

Assuming then for instance $M_{\rm PBH}=10^2$~M$_\odot$ and $c_{\rm II}=23$~km/s (which corresponds to a temperature $T_{\rm II}=4\times10^4$ K~\cite{Park:2012cr}) the various velocities take the form displayed in the left panel of Fig.~\ref{fig: vel_lambda}. The dashed and dotted lines show the two critical velocities\footnote{The two critical velocities cross at around $z\simeq 4.4\times10^{4}$, which can be considered as the redshift above which the PR13 model breaks down. This is safely above the redshifts relevant for this analysis.}, while the blue and red curves show the sound speed of the baryons and the PBH proper velocity, respectively. From the figure it is clear that in a realistic cosmological setup the system is always in the intermediate velocity regime below $z\sim\mathcal{O}(10^4)$, which are the redshifts of interest for the CMB constraints \cite{Slatyer2016General}.

The corresponding values of $\lambda$ in the PR13 model are shown in blue in the right panel of Fig. \ref{fig: vel_lambda} as a function of redshift. For reference, the Bondi (i.e., $\lambda=1$) and BHL limits are shown as dashed and dotted lines, respectively. For further comparison, the values derived in~\cite{AliHaimoud2017Cosmic} (AHK17) for the spherical accretion scenario and in~\cite{Poulin2017CMB} (PSCCK17) for the disk accretion scenario are also shown in orange and red, respectively. As expected, the PR13 model introduces a large suppression of the accretion rate with respect to all other models for $z<\mathcal{O}(10^3)$, while it approaches $\lambda_{\rm BHL}$ at higher redshifts since $v_\infty$ approaches $v_{\rm R}$ (as shown in the left panel).

\subsection{Emission efficiency}\label{sec: eff}
Once the accretion rate has been determined, it is possible to parameterize the corresponding luminosity of the system as
\begin{equation}
    L = \epsilon  \dot M_{\rm PBH} \, ,
\end{equation}
where $\epsilon$ represents the emission efficiency. In PR13 (see Eq. (12) of \cite{Sugimura:2020rdw}), $\epsilon$ has been defined with the phenomenological form
\begin{equation}
    \epsilon = \epsilon_0 \, {\rm min}\left( 1, \frac{\dot M_{\rm PBH}}{L_{\rm Edd}}\right)
\end{equation}
where $L_{\rm edd}$ is the Eddington luminosity and $\epsilon_0$ has been fixed to the benchmark value $\epsilon_0=0.1$. The analytic results of \cite{AliHaimoud2017Cosmic}, however, point towards values of $\epsilon_0$ that are orders of magnitude lower. In particular within the photo-ionization scenario discussed in \cite{AliHaimoud2017Cosmic} (which is the physically closest one to the PR13 model with a sharp I-front), from Fig. 6 of the reference one can infer that after recombination
\begin{align}
    \frac{\epsilon}{\dot{m}}=\frac{L}{\dot{M}_{\rm PBH}}\frac{L_{\rm edd}}{\dot{M}_{\rm PBH}}\simeq 10^{-3} \Leftrightarrow L \simeq 10^{-3} \frac{\dot{M}_{\rm PBH}^2}{L_{\rm edd}}\,,
\end{align}
implying that $\epsilon_0\simeq 10^{-3}$ independenlty of the PBH mass (the collisional ionization scenario would predict an even lower $\epsilon_0$ value). We will therefore consider both values of $\epsilon_0$ as benchmark values for the determination of the final luminosity, considering however the one inferred from \cite{AliHaimoud2017Cosmic} as more conservative and reliable. 
\begin{figure}[t]
    \centering
    \includegraphics[width=0.92\columnwidth]{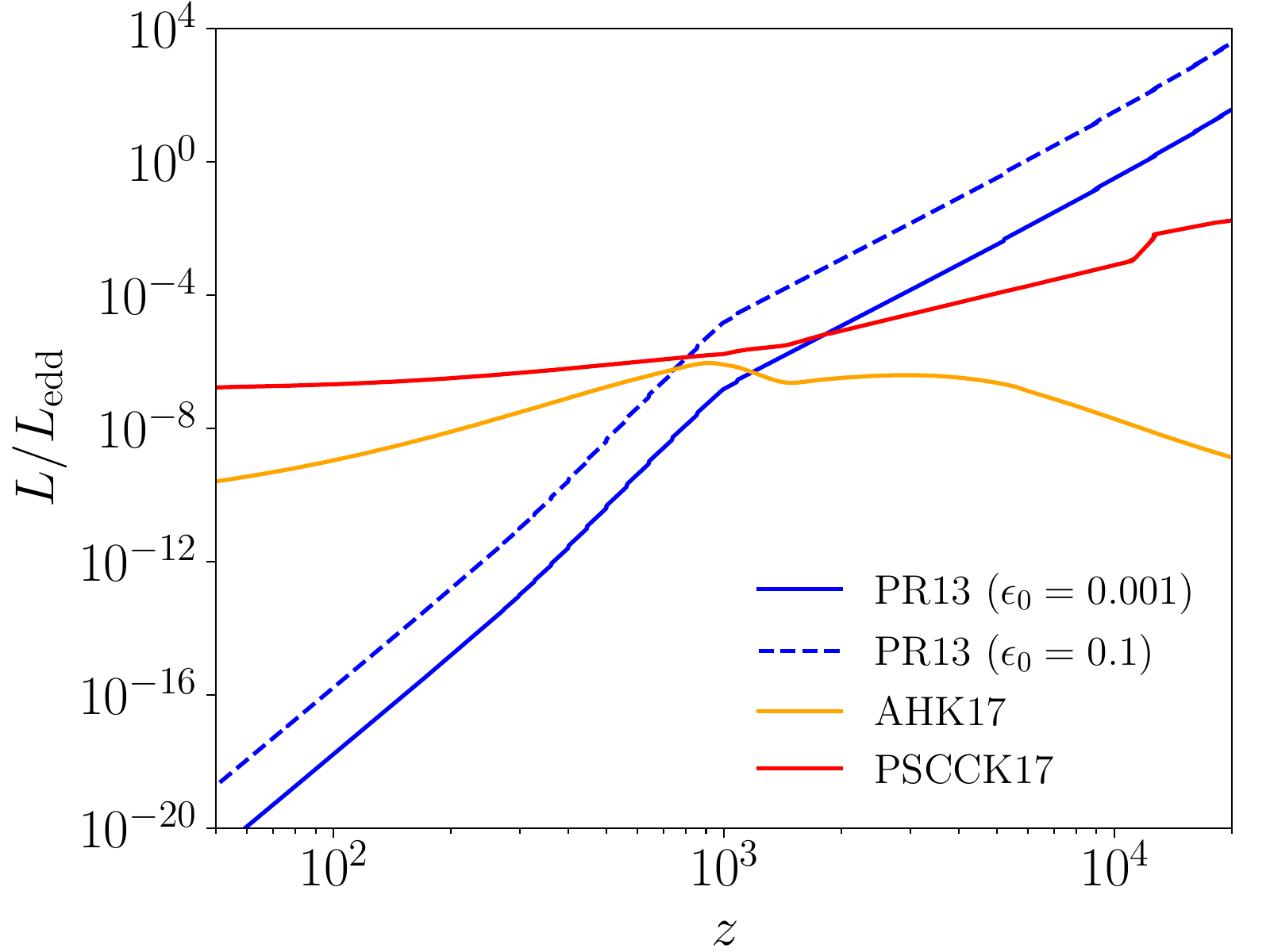}
    \caption{Same as in the right panel of Fig. \ref{fig: vel_lambda} but showing the luminosity of the system normalized to the Eddington luminosity $L_{\rm edd}$. For the fraction of energy in ions $\delta$ in the disk accretion scenario we assume $\delta=0.1$, as done in~\cite{Poulin2017CMB}.}
    \label{fig: L_over_Ledd}
\end{figure}

The corresponding (averaged\footnote{For the calculation of the luminosity (and of the quantities that depend upon it) we always average over the PBH velocity as discussed in Sec. IIF of \cite{AliHaimoud2017Cosmic} and following the implementation of~\cite{Piga:2022ysp}. Not doing so biases the results towards lower luminosities by a factor $5-50$.}) luminosities are shown in Fig. \ref{fig: L_over_Ledd} (solid and dashed blue lines) compared to the luminosity predicted within the spherical and disk accretion scenarios of \cite{AliHaimoud2017Cosmic} and \cite{Poulin2017CMB} for the same PBH mass. Note that, while in the spherical accretion scenario we can employ the photo-ionization limit, which is more realistic in the context of the PR13 model, this is not possible in the disk accretion scenario which has only been developed for the collisional ionization case. Therefore, all curves shown for the disk accretion case are to be taken as a lower limit on what the corresponding photo-ionization case would predict. 

One sees in Fig.~\ref{fig: L_over_Ledd} that all the predictions are relatively comparable around the time of recombination, while the PR13 model predicts luminosities orders of magnitude lower than in the other scenarios at lower redshifts. This means that, since the redshifts around $z\sim \mathcal{O}(500)$ are the ones CMB anisotropies are the most sensitive to in the context of energy injections \cite{Slatyer2015IndirectI, Slatyer2016General}, the CMB bounds derived within the PR13 scenario are going to be relaxed with respect to the other models. 

We conclude this discussion on the luminosity by pointing out that the PR13 model assumes an accretion radius (i.e., the equivalent of the Bondi radius), $M_{\rm PBH}/(v_{\rm II}^2 + c_{\rm II}^2)$, contained inside the ionized region. The size of the latter is fixed by the competition between the amount of ionizing radiation produced (set by the luminosity) and either the recombination rate or the neutral gas inflow through the I-front. Following the prescription of~\cite{Sugimura:2020rdw} we checked that, even for $\epsilon_0 = 0.001$, this holds true as long as $M_{\rm PBH} \ge 1~{\rm M_\odot}$ and $z \gtrsim 200$.

\subsection{Impact on the CMB}\label{sec: imp}
In light of these results, it is possible to calculate the energy injection rate
\begin{align}
    \left.\frac{\text{d}E}{\text{d}t\text{d}V}\right|_{\text{inj}}  = \rho_{\rm cdm}\, f_{\rm PBH}\, \frac{L}{M_{\rm PBH}}
\end{align}
and the corresponding energy deposition rate
\begin{align}
    \left.\frac{\text{d}E}{\text{d}t\text{d}V}\right|_{\text{dep}, c} = \left.\frac{\text{d}E}{\text{d}t\text{d}V}\right|_{\text{inj}} f_{\rm eff}\, \chi_c\,,
\end{align}
where~$f_{\rm eff}$ and~$\chi_c$ are the deposition efficiency and deposition fraction per channel, respectively (see e.g., Sec.~2.4.1 of \cite{Lucca2019Synergy} for a more in-depth discussion with the same notation). For the calculation of $f_{\rm eff}$ we follow section~IV of \cite{AliHaimoud2017Cosmic} (as implemented by \cite{Piga:2022ysp} following the previous HYREC implementation \cite{AliHaimoud2010HyRec, Lee2020HYREC}), while $\chi_c$ is computed according to table~V of \cite{Galli2013Systematic}.

\begin{figure}[t]
    \centering
    \includegraphics[width=\columnwidth]{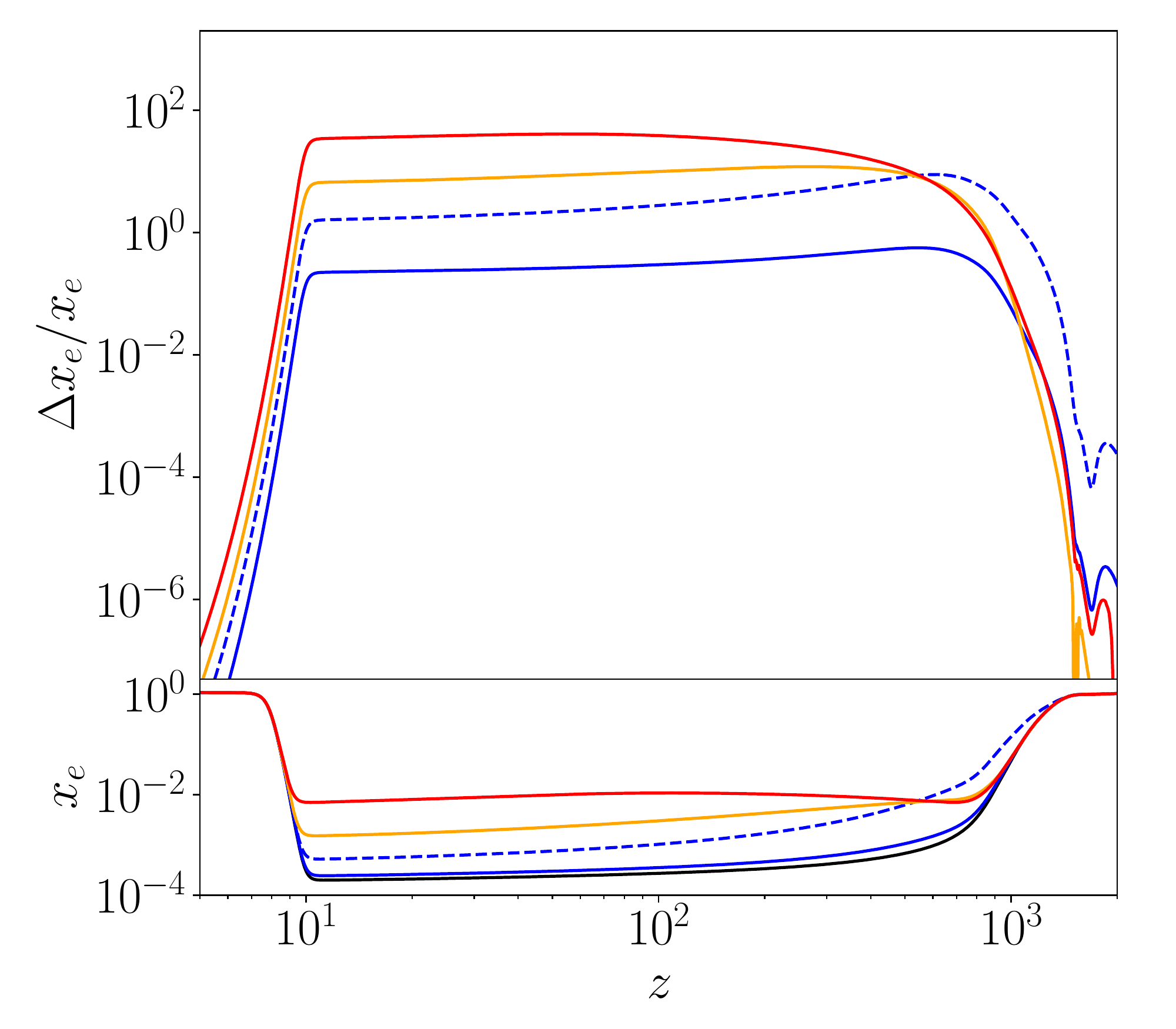}
    \caption{Impact of the accretion models shown in Fig. \ref{fig: L_over_Ledd} on the free electron fraction $x_e$.}
    \label{fig: xe}
\end{figure}

\begin{figure}[t]
    \centering
    \includegraphics[width=0.98\columnwidth]{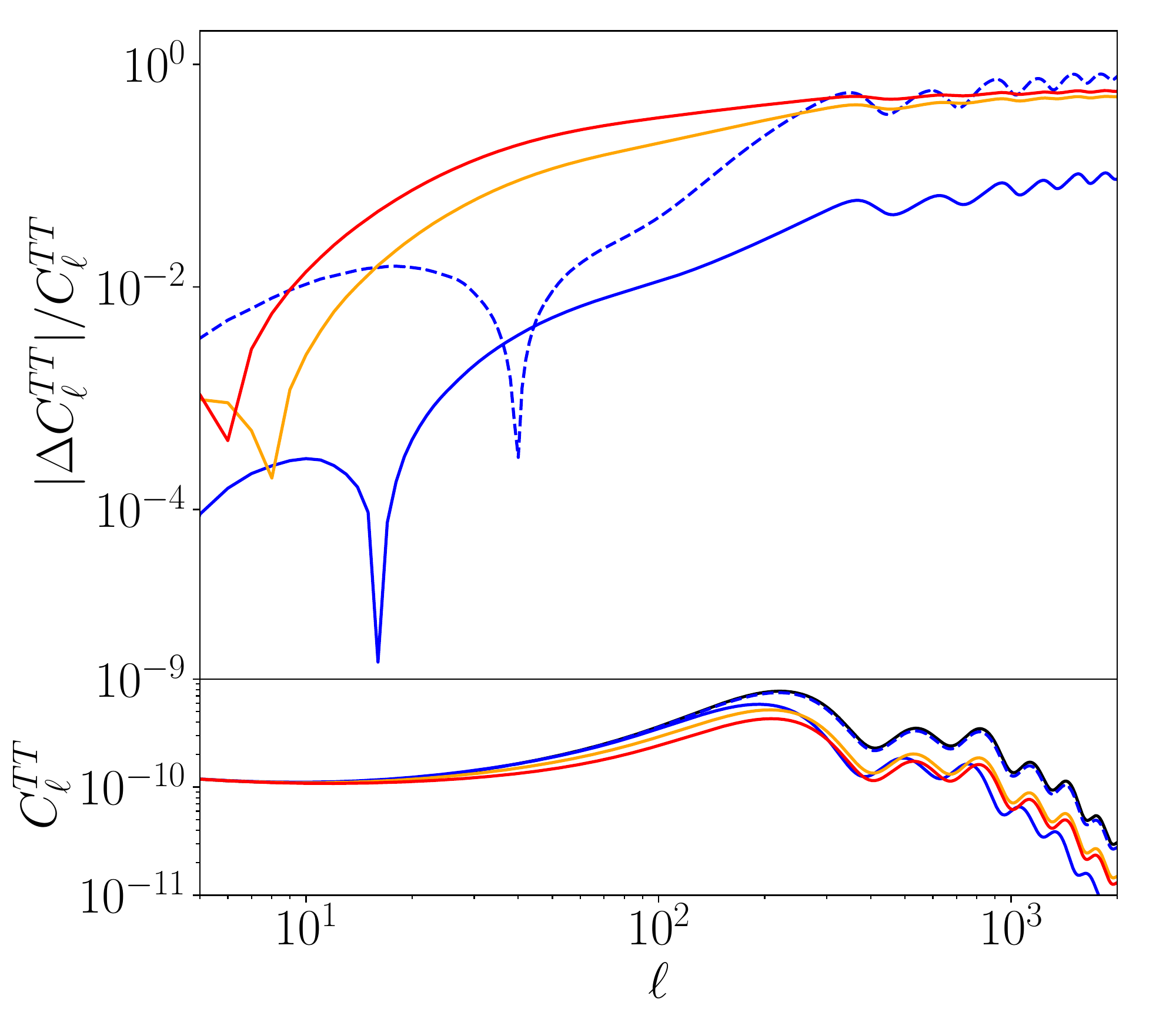} \\
    \includegraphics[width=0.98\columnwidth]{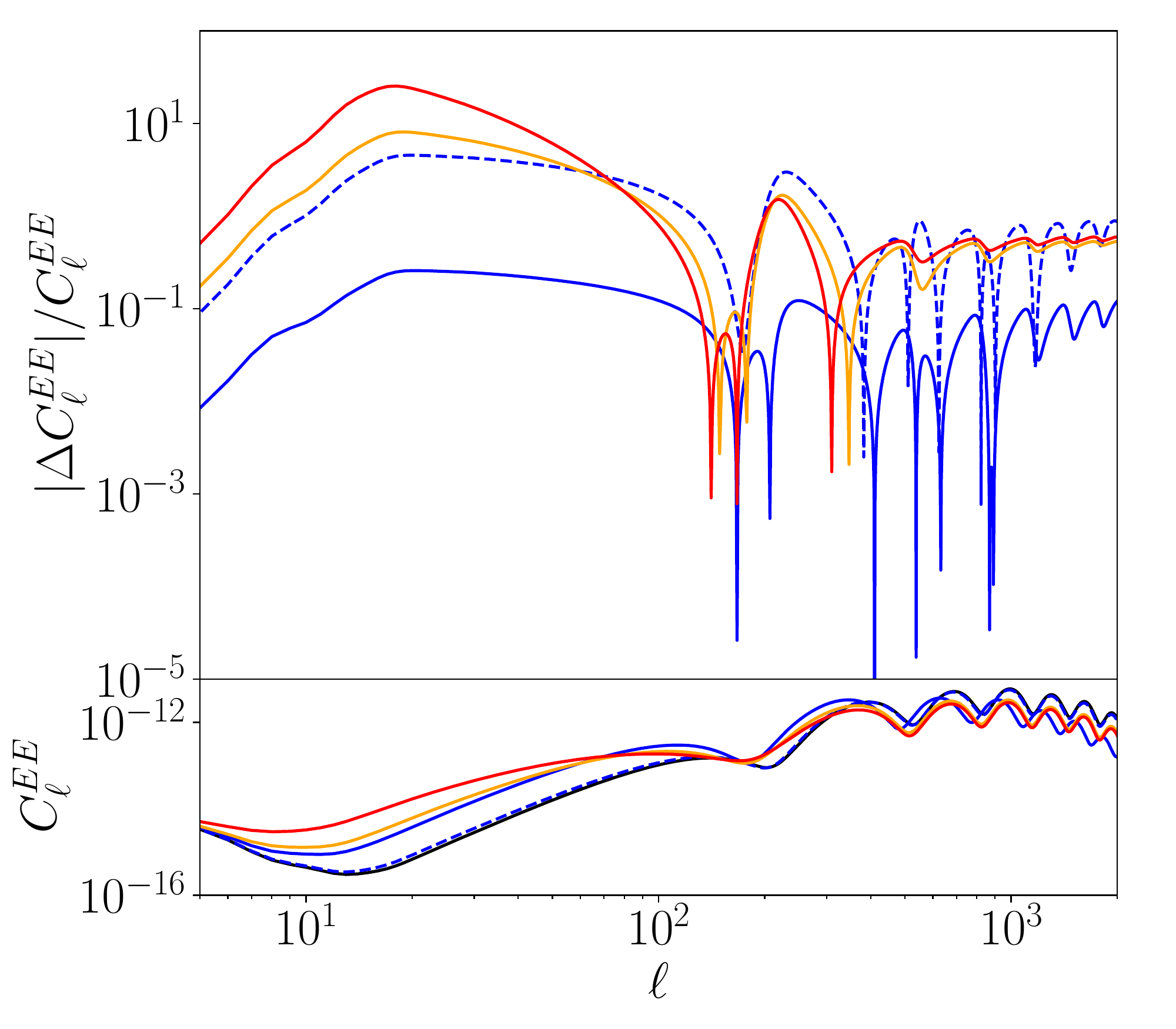}
    \caption{Impact of the accretion models shown in Fig.~\ref{fig: L_over_Ledd} on the CMB temperature (top) and polarization (bottom) anisotropy power spectra.}
    \label{fig: Cl}
\end{figure}

Most notably, once deposited the emitted radiation can partly ionize the neutral universe and modify the free electron fraction $x_e$. The cases corresponding to those shown in Fig. \ref{fig: L_over_Ledd} are displayed in Fig. \ref{fig: xe} for $f_{\rm PBH}=1$. As expected, also in this case we observe a suppressed impact of the emission from the accretion in the PR13 model with respect to the other scenarios. This behavior is directly translated in terms of the CMB anisotropy power spectra, which we show in Fig.~\ref{fig: Cl}. 

\section{Numerical setup}\label{sec: num}
For the calculation of the physical quantities displayed in Figs.\ref{fig: vel_lambda}-\ref{fig: Cl} we use the latest version of the cosmological Boltzmann solver \texttt{CLASS} \cite{Lesgourgues2011CosmicI, Blas2011Cosmic}. Specifically, we rely on the energy injection implementation discussed in~\cite{Lucca2019Synergy} (in turn largely based on the \texttt{ExoCLASS} extension of \texttt{CLASS}~\cite{Stocker2018Exotic}), which we have extended to include the PR13 model discussed in the previous section. This modified \texttt{CLASS} version is then used in interface with the parameter extraction code \texttt{MontePython}~\cite{Audren2013Conservative, Brinckmann2018MontePython} to derive the cosmological constraints on the PBH abundance presented in the following section. Specifically, we scanned the 6+1 parameter space
\begin{align}
    \{\omega_b\,, \omega_{\rm cdm}\,, h\,, A_s\,, n_s\,, z_{\rm reio}\}+f_{\rm PBH}
\end{align}
with a series of Markov Chain Monte Carlo (MCMC) runs for fixed values of the PBH mass in the range $[10-10^4]$ M$_{\odot}$.\footnote{Here $\omega_b$ and $\omega_{\rm cdm}$ are the baryon and DM energy densities, respectively, $h$ is the dimensionless Hubble rate, $A_s$ and $n_s$ are the amplitude and the spectral index of the primordial power spectrum and $z_{\rm reio}$ is the reionization redshift.} These cosmological parameters have then been constrained with temperature, polarization and lensing data gathered by the Planck mission \cite{Aghanim2018PlanckVI}, specifically using the 2018 high-$\ell$ TTTEEE, low-$\ell$ EE, low-$\ell$ TT and lensing likelihoods \cite{Aghanim2018PlanckV}. The runs have been considered converged with the Gelman-Rubin criterion $|R-1|<0.03$~\cite{Gelman1992Inference}.

\section{Results}\label{sec: res}
\begin{figure}[t]
    \centering
    \includegraphics[width=\columnwidth]{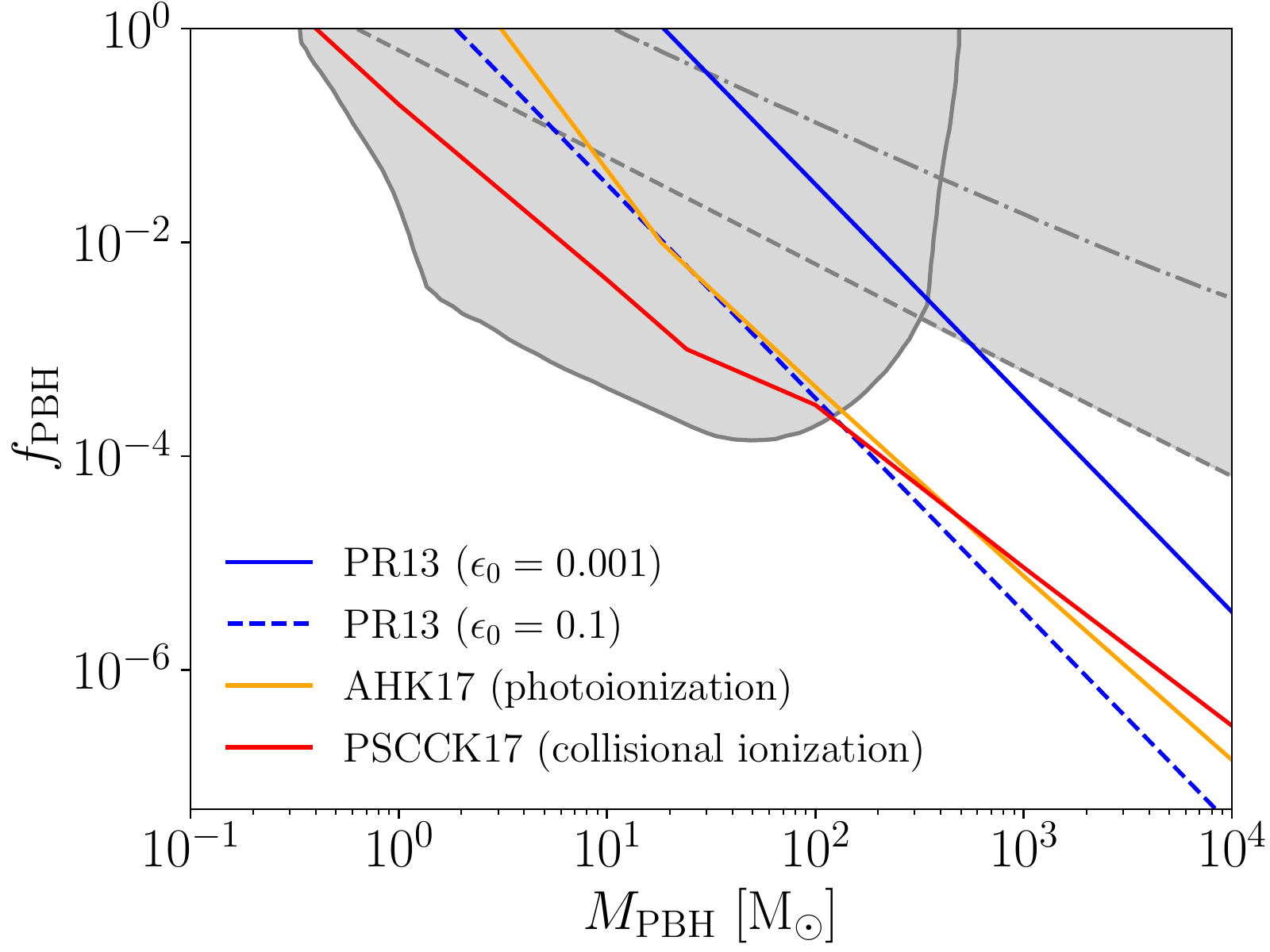}
    \caption{95\% CL bounds on the PBH abundance for all models discussed in the previous section. The gray area highlights the region of parameter space excluded by complementary observations (from the merger rates of binary BHs inferred by LIGO/Virgo \cite{Hutsi:2020sol} -- solid -- and from ultra-faint dwarf galaxies~\cite{Brandt:2016aco} -- Segues I in dashed and Eridanus II in dashed-dotted).}
    \label{fig: bounds}
\end{figure}

The final 95\% CL bounds on the PBH abundance (derived from the MCMCs according to Sec. \ref{sec: num}) are shown in Fig.~\ref{fig: bounds} for all models discussed in the previous section.\footnote{We find no statistically significant deviation from the standard Planck values reported in \cite{Aghanim2018PlanckVI} for the other cosmological parameters.} Since, according to Sec. \ref{sec: eff}, $L\propto \epsilon_0 M_{\rm PBH}^3$ in the PR13 model (neglecting the weak dependence of $c_\infty$ on the modified matter temperature evolution) for the total injected energy we have that
\begin{align}
    \frac{\text{d}E}{\text{d}t\text{d}E} \propto f_{\rm PBH} \frac{L}{M_{\rm PBH}} \propto f_{\rm PBH} \epsilon_0 M^2_{\rm PBH}\,.
\end{align}
This means that for the final bounds, where the value of $\epsilon_0$ is fixed, we can expect a direct dependence of the form $f_{\rm PBH}\propto M_{\rm PBH}^{-2}$, behavior that the PR13 constraints perfectly match in the figure. At the same time, this also implies that the bounds on the abundance $f_{\rm PBH}$ obtained for a given value of $\epsilon_0$ can be simply rescaled for any other value of this quantity without the need for additional simulations.

Moreover, we note that under the assumption of very efficient emissions ($\epsilon_0=0.1$, dashed blue line) the predicted CMB constraints are almost identical to those derived in the context of the spherical (assuming photo-ionization, orange line) and disk (assuming collisional ionization, red line) accretion scenarios described in \cite{AliHaimoud2017Cosmic} and \cite{Poulin2017CMB}. However, for more conservative values of $\epsilon_0$ ($\epsilon_0=0.001$, solid blue line) the bounds are suppressed by two or more orders of magnitude, allowing for PBH masses as high as 20 M$_\odot$ for the case where PBHs make up for the full DM content of the universe. In comparison, $f_{\rm PBH }=1$ was excluded for masses higher than approximately $2-3$ M$_\odot$ in the spherical accretion scenario considered in~\cite{AliHaimoud2017Cosmic} as well as in the less conservative $\epsilon_0=0.1$ case. Even smaller masses are excluded in the disk accretion scenario, in particular considering that the shown constraints have been derived assuming collisional ionization, while the more efficient (and hence more constraining) photo-ionization limit would need to be considered to be consistent with the presence of an I-front. In other words, the PR13 model with realistic emission efficiencies might open up the allowed PBH mass range (for any value of $f_{\rm PBH}$) by more than one order of magnitude with respect to previous proposals.

In Fig.~\ref{fig: bounds} we also compare the aforementioned CMB bounds with the cumulative constraints coming from the ultra-faint dwarf galaxies Segues~I and Eridanus II \cite{Brandt:2016aco} and from the merger rates of binary BHs inferred by LIGO-Virgo \cite{Hutsi:2020sol}. This shows that the CMB bounds in the conservative limit of the PR13 model are only dominant for masses larger than approximately 500 M$_\odot$, freeing a significant portion of parameter space above 100~M$_\odot$. It is furthermore interesting to point out that the scaling of the emission luminosity shown in Fig.~\ref{fig: L_over_Ledd} suggests that all late-time bounds on PBH accretion, such as the ones derived in~\cite{Ziparo:2022fnc} from the cosmic X-ray background radiation and in \cite{Mena:2019nhm} for up-coming 21cm observations, might be even more significantly suppressed in the PR13 model than the CMB bounds discussed here. At the same time, one also needs to consider the fact that late-time effects such as the formation of DM halos might affect the accretion rate, potentially leading to an overall strengthening of the constraints \cite{Serpico2020Cosmic} (see also \cite{DeLuca2020Constraints} for complementary effects weakening the final bounds). A dedicated analysis is however left for future work, considering in particular that the PR13 model as described in Sec. \ref{sec: PR} might not be valid at such low redshifts.

As a final remark, given the very simple mass dependence of the constraints discussed above in the context of the PR13 model, it is straightforward to derive the corresponding limits for an extended normalized PBH mass distribution $f(M_{\rm PBH})$ by calculating the equivalent mass $M_{\rm eq} \equiv \langle f(M_{\rm PBH}) M_{\rm PBH}^2 \rangle^{1/2}$ from which one can extract the corresponding limit on $f_{\rm PBH}$. Here $\langle . \rangle$ denotes the average with respect to the PBH number density, which gives different expressions for different definitions and normalizations of the PBH mass distribution, see Table 1 of~\cite{Escriva:2022bwe}.  As an illustrative example, we find that $M_{\rm eq} = 5.0$~M$_\odot $ for the extended mass distribution used~\cite{Escriva:2022bwe} obtained for an almost scale-invariant primordial power spectrum with a spectral index $n_{\rm s} = 0.975$ and including the imprints from the QCD transition on the PBH formation.  As one can see from Fig.~\ref{fig: bounds}, for such a value of $M_{\rm eq} $ our CMB limits still allow $f_{\rm PBH} = 1$ for the conservative case $\epsilon_0 = 10^{-3}$ and only impose that $f_{\rm PBH} \simeq 0.1$ for $\epsilon_0 = 0.1$. This class of mass distributions, even with a larger value of $n_{\rm s}$ or with a negative running, is therefore not constrained by CMB anisotropies once the effect of ionization fronts is taken into account.

Overall, we find that our relaxed CMB limits re-open the possibility to explain some of the BH merger events observed by LVK with a PBH population, in particular in the case of GW190521 which has been associated to at least one BH in the pair-instability mass gap that would require $f_{\rm PBH} \simeq 10^{-3}-10^{-4}$ for a peaked mass distribution \cite{DeLuca:2020sae}. This conclusion would only be reinforced under the assumption of more realistic PBH mass distributions, as argued above.

\section{Summary and conclusions}\label{sec: conc} 
PBHs have been recently brought forward as a viable explanation for at least some of the GW events observed by LVK if their mass lays in the approximate range between $10-100$~M$_\odot$. Nevertheless, this region of the parameter space is also tightly constrained by the imprints of PBH accretion on CMB anisotropies. In order to test the compatibility of these astrophysical hints and cosmological constraints it is therefore fundamental to develop a model of PBH accretion as realistic, and yet conservative, as possible.

In this work we consider a model developed in \cite{Park:2012cr} (PR13) which accurately accounts for the physics related to the presence of ionization fronts (I-fronts) preceding the PBHs in their motion. Importantly, its validity has been verified on the basis of numerical simulations \cite{Sugimura:2020rdw}. Building on the work carried out by \cite{Park:2012cr, Sugimura:2020rdw}, here we show that in a cosmological context the presence of such I-fronts always leads to the formation of a dense shell ahead of the PBH (D-type front) that decreases the accretion rate and hence reduces the radiation emission.

With this model, we obtain more realistic and conservative bounds on the possible fraction of DM that can be made of PBHs, which relax by more than one order of magnitude those derived in previous works. As a result, we find that LVK BH merger rates can be obtained with PBHs without imposing tight restrictions on the shape of their mass distributions. More generally, our work emphasizes that BH accretion is a complex and delicate problem, and that accretion-based limits on PBHs -- like other limits -- are still subject to large uncertainties. In this context, strong claims on the ability of PBHs to explain the DM abundance, GW observations and  the existence of supermassive BHs must be cautious and are probably still premature.  

\section*{Acknowledgements}
The authors thank Massimo Ricotti and Francesca Scarcella for the useful feedback on the manuscript. GF acknowledges support of the ARC program of the Federation Wallonie-Bruxelles and of the Excellence of Science (EoS) project No. 30820817 - be.h “The H boson gateway to physics beyond the Standard Model”. ML is supported by an F.R.S.-FNRS fellowship. SC acknowledges support from a Start-up Grant of the Belgian Francqui Foundation and a Mandat d'Inpulsion Scientifique (MIS) from the Belgian Fund for Research F.R.S.-FNRS. Computational resources have been provided by the Consortium des \'Equipements de Calcul Intensif (C\'ECI), funded by the Fonds de la Recherche Scientifique de Belgique (F.R.S.-FNRS) under Grant No. 2.5020.11 and by the Walloon Region.

\bibliography{bibliography}{}

\begin{thebibliography}{57}%
\makeatletter
\providecommand \@ifxundefined [1]{%
 \@ifx{#1\undefined}
}%
\providecommand \@ifnum [1]{%
 \ifnum #1\expandafter \@firstoftwo
 \else \expandafter \@secondoftwo
 \fi
}%
\providecommand \@ifx [1]{%
 \ifx #1\expandafter \@firstoftwo
 \else \expandafter \@secondoftwo
 \fi
}%
\providecommand \natexlab [1]{#1}%
\providecommand \enquote  [1]{``#1''}%
\providecommand \bibnamefont  [1]{#1}%
\providecommand \bibfnamefont [1]{#1}%
\providecommand \citenamefont [1]{#1}%
\providecommand \href@noop [0]{\@secondoftwo}%
\providecommand \href [0]{\begingroup \@sanitize@url \@href}%
\providecommand \@href[1]{\@@startlink{#1}\@@href}%
\providecommand \@@href[1]{\endgroup#1\@@endlink}%
\providecommand \@sanitize@url [0]{\catcode `\\12\catcode `\$12\catcode
  `\&12\catcode `\#12\catcode `\^12\catcode `\_12\catcode `\%12\relax}%
\providecommand \@@startlink[1]{}%
\providecommand \@@endlink[0]{}%
\providecommand \url  [0]{\begingroup\@sanitize@url \@url }%
\providecommand \@url [1]{\endgroup\@href {#1}{\urlprefix }}%
\providecommand \urlprefix  [0]{URL }%
\providecommand \Eprint [0]{\href }%
\providecommand \doibase [0]{https://doi.org/}%
\providecommand \selectlanguage [0]{\@gobble}%
\providecommand \bibinfo  [0]{\@secondoftwo}%
\providecommand \bibfield  [0]{\@secondoftwo}%
\providecommand \translation [1]{[#1]}%
\providecommand \BibitemOpen [0]{}%
\providecommand \bibitemStop [0]{}%
\providecommand \bibitemNoStop [0]{.\EOS\space}%
\providecommand \EOS [0]{\spacefactor3000\relax}%
\providecommand \BibitemShut  [1]{\csname bibitem#1\endcsname}%
\let\auto@bib@innerbib\@empty
\bibitem [{\citenamefont {Carr}\ \emph {et~al.}(2016)\citenamefont {Carr},
  \citenamefont {Kuhnel},\ and\ \citenamefont {Sandstad}}]{Carr2016Primordial}%
  \BibitemOpen
  \bibfield  {author} {\bibinfo {author} {\bibfnamefont {B.}~\bibnamefont
  {Carr}}, \bibinfo {author} {\bibfnamefont {F.}~\bibnamefont {Kuhnel}},\ and\
  \bibinfo {author} {\bibfnamefont {M.}~\bibnamefont {Sandstad}},\ }\bibfield
  {title} {\bibinfo {title} {{Primordial Black Holes as Dark Matter}},\ }\href
  {https://doi.org/10.1103/PhysRevD.94.083504} {\bibfield  {journal} {\bibinfo
  {journal} {Phys. Rev.}\ }\textbf {\bibinfo {volume} {D94}},\ \bibinfo {pages}
  {083504} (\bibinfo {year} {2016})},\ \Eprint
  {https://arxiv.org/abs/1607.06077} {arXiv:1607.06077 [astro-ph.CO]}
  \BibitemShut {NoStop}%
\bibitem [{\citenamefont {Carr}\ and\ \citenamefont
  {Kuhnel}(2020)}]{Carr2020Primordial}%
  \BibitemOpen
  \bibfield  {author} {\bibinfo {author} {\bibfnamefont {B.}~\bibnamefont
  {Carr}}\ and\ \bibinfo {author} {\bibfnamefont {F.}~\bibnamefont {Kuhnel}},\
  }\bibfield  {title} {\bibinfo {title} {{Primordial Black Holes as Dark
  Matter: Recent Developments}},\ }\href
  {https://doi.org/10.1146/annurev-nucl-050520-125911} {\bibfield  {journal}
  {\bibinfo  {journal} {Ann. Rev. Nucl. Part. Sci.}\ }\textbf {\bibinfo
  {volume} {70}},\ \bibinfo {pages} {355} (\bibinfo {year} {2020})},\ \Eprint
  {https://arxiv.org/abs/2006.02838} {arXiv:2006.02838 [astro-ph.CO]}
  \BibitemShut {NoStop}%
\bibitem [{\citenamefont {Abbott}\ \emph
  {et~al.}(2021{\natexlab{a}})\citenamefont {Abbott} \emph
  {et~al.}}]{LIGOScientific:2021djp}%
  \BibitemOpen
  \bibfield  {author} {\bibinfo {author} {\bibfnamefont {R.}~\bibnamefont
  {Abbott}} \emph {et~al.} (\bibinfo {collaboration} {LIGO Scientific, VIRGO,
  KAGRA}),\ }\bibfield  {title} {\bibinfo {title} {{GWTC-3: Compact Binary
  Coalescences Observed by LIGO and Virgo During the Second Part of the Third
  Observing Run}},\ }\href@noop {} {\  (\bibinfo {year}
  {2021}{\natexlab{a}})},\ \Eprint {https://arxiv.org/abs/2111.03606}
  {arXiv:2111.03606 [gr-qc]} \BibitemShut {NoStop}%
\bibitem [{\citenamefont {Abbott}\ \emph
  {et~al.}(2021{\natexlab{b}})\citenamefont {Abbott} \emph
  {et~al.}}]{LIGOScientific:2021usb}%
  \BibitemOpen
  \bibfield  {author} {\bibinfo {author} {\bibfnamefont {R.}~\bibnamefont
  {Abbott}} \emph {et~al.} (\bibinfo {collaboration} {LIGO Scientific,
  VIRGO}),\ }\bibfield  {title} {\bibinfo {title} {{GWTC-2.1: Deep Extended
  Catalog of Compact Binary Coalescences Observed by LIGO and Virgo During the
  First Half of the Third Observing Run}},\ }\href@noop {} {\  (\bibinfo {year}
  {2021}{\natexlab{b}})},\ \Eprint {https://arxiv.org/abs/2108.01045}
  {arXiv:2108.01045 [gr-qc]} \BibitemShut {NoStop}%
\bibitem [{\citenamefont {Abbott}\ \emph
  {et~al.}(2021{\natexlab{c}})\citenamefont {Abbott} \emph
  {et~al.}}]{LIGOScientific:2020ibl}%
  \BibitemOpen
  \bibfield  {author} {\bibinfo {author} {\bibfnamefont {R.}~\bibnamefont
  {Abbott}} \emph {et~al.} (\bibinfo {collaboration} {LIGO Scientific,
  Virgo}),\ }\bibfield  {title} {\bibinfo {title} {{GWTC-2: Compact Binary
  Coalescences Observed by LIGO and Virgo During the First Half of the Third
  Observing Run}},\ }\href {https://doi.org/10.1103/PhysRevX.11.021053}
  {\bibfield  {journal} {\bibinfo  {journal} {Phys. Rev. X}\ }\textbf {\bibinfo
  {volume} {11}},\ \bibinfo {pages} {021053} (\bibinfo {year}
  {2021}{\natexlab{c}})},\ \Eprint {https://arxiv.org/abs/2010.14527}
  {arXiv:2010.14527 [gr-qc]} \BibitemShut {NoStop}%
\bibitem [{\citenamefont {Abbott}\ \emph {et~al.}(2019)\citenamefont {Abbott}
  \emph {et~al.}}]{LIGOScientific:2018mvr}%
  \BibitemOpen
  \bibfield  {author} {\bibinfo {author} {\bibfnamefont {B.~P.}\ \bibnamefont
  {Abbott}} \emph {et~al.} (\bibinfo {collaboration} {LIGO Scientific,
  Virgo}),\ }\bibfield  {title} {\bibinfo {title} {{GWTC-1: A
  Gravitational-Wave Transient Catalog of Compact Binary Mergers Observed by
  LIGO and Virgo during the First and Second Observing Runs}},\ }\href
  {https://doi.org/10.1103/PhysRevX.9.031040} {\bibfield  {journal} {\bibinfo
  {journal} {Phys. Rev. X}\ }\textbf {\bibinfo {volume} {9}},\ \bibinfo {pages}
  {031040} (\bibinfo {year} {2019})},\ \Eprint
  {https://arxiv.org/abs/1811.12907} {arXiv:1811.12907 [astro-ph.HE]}
  \BibitemShut {NoStop}%
\bibitem [{\citenamefont {Abbott}\ \emph
  {et~al.}(2020{\natexlab{a}})\citenamefont {Abbott} \emph
  {et~al.}}]{LIGOScientific:2020iuh}%
  \BibitemOpen
  \bibfield  {author} {\bibinfo {author} {\bibfnamefont {R.}~\bibnamefont
  {Abbott}} \emph {et~al.} (\bibinfo {collaboration} {LIGO Scientific,
  Virgo}),\ }\bibfield  {title} {\bibinfo {title} {{GW190521: A Binary Black
  Hole Merger with a Total Mass of $150 M_{\odot}$}},\ }\href
  {https://doi.org/10.1103/PhysRevLett.125.101102} {\bibfield  {journal}
  {\bibinfo  {journal} {Phys. Rev. Lett.}\ }\textbf {\bibinfo {volume} {125}},\
  \bibinfo {pages} {101102} (\bibinfo {year} {2020}{\natexlab{a}})},\ \Eprint
  {https://arxiv.org/abs/2009.01075} {arXiv:2009.01075 [gr-qc]} \BibitemShut
  {NoStop}%
\bibitem [{\citenamefont {Abbott}\ \emph
  {et~al.}(2020{\natexlab{b}})\citenamefont {Abbott} \emph
  {et~al.}}]{LIGOScientific:2020zkf}%
  \BibitemOpen
  \bibfield  {author} {\bibinfo {author} {\bibfnamefont {R.}~\bibnamefont
  {Abbott}} \emph {et~al.} (\bibinfo {collaboration} {LIGO Scientific,
  Virgo}),\ }\bibfield  {title} {\bibinfo {title} {{GW190814: Gravitational
  Waves from the Coalescence of a 23 Solar Mass Black Hole with a 2.6 Solar
  Mass Compact Object}},\ }\href {https://doi.org/10.3847/2041-8213/ab960f}
  {\bibfield  {journal} {\bibinfo  {journal} {Astrophys. J. Lett.}\ }\textbf
  {\bibinfo {volume} {896}},\ \bibinfo {pages} {L44} (\bibinfo {year}
  {2020}{\natexlab{b}})},\ \Eprint {https://arxiv.org/abs/2006.12611}
  {arXiv:2006.12611 [astro-ph.HE]} \BibitemShut {NoStop}%
\bibitem [{\citenamefont {Bird}\ \emph {et~al.}(2016)\citenamefont {Bird},
  \citenamefont {Cholis}, \citenamefont {Mu\~noz}, \citenamefont
  {Ali-Ha\"\i{}moud}, \citenamefont {Kamionkowski}, \citenamefont {Kovetz},
  \citenamefont {Raccanelli},\ and\ \citenamefont {Riess}}]{Bird:2016dcv}%
  \BibitemOpen
  \bibfield  {author} {\bibinfo {author} {\bibfnamefont {S.}~\bibnamefont
  {Bird}}, \bibinfo {author} {\bibfnamefont {I.}~\bibnamefont {Cholis}},
  \bibinfo {author} {\bibfnamefont {J.~B.}\ \bibnamefont {Mu\~noz}}, \bibinfo
  {author} {\bibfnamefont {Y.}~\bibnamefont {Ali-Ha\"\i{}moud}}, \bibinfo
  {author} {\bibfnamefont {M.}~\bibnamefont {Kamionkowski}}, \bibinfo {author}
  {\bibfnamefont {E.~D.}\ \bibnamefont {Kovetz}}, \bibinfo {author}
  {\bibfnamefont {A.}~\bibnamefont {Raccanelli}},\ and\ \bibinfo {author}
  {\bibfnamefont {A.~G.}\ \bibnamefont {Riess}},\ }\bibfield  {title} {\bibinfo
  {title} {{Did LIGO detect dark matter?}},\ }\href
  {https://doi.org/10.1103/PhysRevLett.116.201301} {\bibfield  {journal}
  {\bibinfo  {journal} {Phys. Rev. Lett.}\ }\textbf {\bibinfo {volume} {116}},\
  \bibinfo {pages} {201301} (\bibinfo {year} {2016})},\ \Eprint
  {https://arxiv.org/abs/1603.00464} {arXiv:1603.00464 [astro-ph.CO]}
  \BibitemShut {NoStop}%
\bibitem [{\citenamefont {Clesse}\ and\ \citenamefont
  {Garc\'\i{}a-Bellido}(2017)}]{Clesse:2016vqa}%
  \BibitemOpen
  \bibfield  {author} {\bibinfo {author} {\bibfnamefont {S.}~\bibnamefont
  {Clesse}}\ and\ \bibinfo {author} {\bibfnamefont {J.}~\bibnamefont
  {Garc\'\i{}a-Bellido}},\ }\bibfield  {title} {\bibinfo {title} {{The
  clustering of massive Primordial Black Holes as Dark Matter: measuring their
  mass distribution with Advanced LIGO}},\ }\href
  {https://doi.org/10.1016/j.dark.2016.10.002} {\bibfield  {journal} {\bibinfo
  {journal} {Phys. Dark Univ.}\ }\textbf {\bibinfo {volume} {15}},\ \bibinfo
  {pages} {142} (\bibinfo {year} {2017})},\ \Eprint
  {https://arxiv.org/abs/1603.05234} {arXiv:1603.05234 [astro-ph.CO]}
  \BibitemShut {NoStop}%
\bibitem [{\citenamefont {Sasaki}\ \emph {et~al.}(2016)\citenamefont {Sasaki},
  \citenamefont {Suyama}, \citenamefont {Tanaka},\ and\ \citenamefont
  {Yokoyama}}]{Sasaki:2016jop}%
  \BibitemOpen
  \bibfield  {author} {\bibinfo {author} {\bibfnamefont {M.}~\bibnamefont
  {Sasaki}}, \bibinfo {author} {\bibfnamefont {T.}~\bibnamefont {Suyama}},
  \bibinfo {author} {\bibfnamefont {T.}~\bibnamefont {Tanaka}},\ and\ \bibinfo
  {author} {\bibfnamefont {S.}~\bibnamefont {Yokoyama}},\ }\bibfield  {title}
  {\bibinfo {title} {{Primordial Black Hole Scenario for the Gravitational-Wave
  Event GW150914}},\ }\href {https://doi.org/10.1103/PhysRevLett.117.061101}
  {\bibfield  {journal} {\bibinfo  {journal} {Phys. Rev. Lett.}\ }\textbf
  {\bibinfo {volume} {117}},\ \bibinfo {pages} {061101} (\bibinfo {year}
  {2016})},\ \bibinfo {note} {[Erratum: Phys.Rev.Lett. 121, 059901 (2018)]},\
  \Eprint {https://arxiv.org/abs/1603.08338} {arXiv:1603.08338 [astro-ph.CO]}
  \BibitemShut {NoStop}%
\bibitem [{\citenamefont {Clesse}\ and\ \citenamefont
  {Garc\'\i{}a-Bellido}(2018)}]{Clesse:2017bsw}%
  \BibitemOpen
  \bibfield  {author} {\bibinfo {author} {\bibfnamefont {S.}~\bibnamefont
  {Clesse}}\ and\ \bibinfo {author} {\bibfnamefont {J.}~\bibnamefont
  {Garc\'\i{}a-Bellido}},\ }\bibfield  {title} {\bibinfo {title} {{Seven Hints
  for Primordial Black Hole Dark Matter}},\ }\href
  {https://doi.org/10.1016/j.dark.2018.08.004} {\bibfield  {journal} {\bibinfo
  {journal} {Phys. Dark Univ.}\ }\textbf {\bibinfo {volume} {22}},\ \bibinfo
  {pages} {137} (\bibinfo {year} {2018})},\ \Eprint
  {https://arxiv.org/abs/1711.10458} {arXiv:1711.10458 [astro-ph.CO]}
  \BibitemShut {NoStop}%
\bibitem [{\citenamefont {Raidal}\ \emph {et~al.}(2017)\citenamefont {Raidal},
  \citenamefont {Vaskonen},\ and\ \citenamefont {Veerm\"ae}}]{Raidal:2017mfl}%
  \BibitemOpen
  \bibfield  {author} {\bibinfo {author} {\bibfnamefont {M.}~\bibnamefont
  {Raidal}}, \bibinfo {author} {\bibfnamefont {V.}~\bibnamefont {Vaskonen}},\
  and\ \bibinfo {author} {\bibfnamefont {H.}~\bibnamefont {Veerm\"ae}},\
  }\bibfield  {title} {\bibinfo {title} {{Gravitational Waves from Primordial
  Black Hole Mergers}},\ }\href {https://doi.org/10.1088/1475-7516/2017/09/037}
  {\bibfield  {journal} {\bibinfo  {journal} {JCAP}\ }\textbf {\bibinfo
  {volume} {09}},\ \bibinfo {pages} {037}},\ \Eprint
  {https://arxiv.org/abs/1707.01480} {arXiv:1707.01480 [astro-ph.CO]}
  \BibitemShut {NoStop}%
\bibitem [{\citenamefont {Raidal}\ \emph {et~al.}(2019)\citenamefont {Raidal},
  \citenamefont {Spethmann}, \citenamefont {Vaskonen},\ and\ \citenamefont
  {Veerm\"ae}}]{Raidal:2018bbj}%
  \BibitemOpen
  \bibfield  {author} {\bibinfo {author} {\bibfnamefont {M.}~\bibnamefont
  {Raidal}}, \bibinfo {author} {\bibfnamefont {C.}~\bibnamefont {Spethmann}},
  \bibinfo {author} {\bibfnamefont {V.}~\bibnamefont {Vaskonen}},\ and\
  \bibinfo {author} {\bibfnamefont {H.}~\bibnamefont {Veerm\"ae}},\ }\bibfield
  {title} {\bibinfo {title} {{Formation and Evolution of Primordial Black Hole
  Binaries in the Early Universe}},\ }\href
  {https://doi.org/10.1088/1475-7516/2019/02/018} {\bibfield  {journal}
  {\bibinfo  {journal} {JCAP}\ }\textbf {\bibinfo {volume} {02}},\ \bibinfo
  {pages} {018}},\ \Eprint {https://arxiv.org/abs/1812.01930} {arXiv:1812.01930
  [astro-ph.CO]} \BibitemShut {NoStop}%
\bibitem [{\citenamefont {Carr}\ \emph {et~al.}(2021)\citenamefont {Carr},
  \citenamefont {Clesse}, \citenamefont {Garc\'\i{}a-Bellido},\ and\
  \citenamefont {K\"uhnel}}]{Carr:2019kxo}%
  \BibitemOpen
  \bibfield  {author} {\bibinfo {author} {\bibfnamefont {B.}~\bibnamefont
  {Carr}}, \bibinfo {author} {\bibfnamefont {S.}~\bibnamefont {Clesse}},
  \bibinfo {author} {\bibfnamefont {J.}~\bibnamefont {Garc\'\i{}a-Bellido}},\
  and\ \bibinfo {author} {\bibfnamefont {F.}~\bibnamefont {K\"uhnel}},\
  }\bibfield  {title} {\bibinfo {title} {{Cosmic conundra explained by thermal
  history and primordial black holes}},\ }\href
  {https://doi.org/10.1016/j.dark.2020.100755} {\bibfield  {journal} {\bibinfo
  {journal} {Phys. Dark Univ.}\ }\textbf {\bibinfo {volume} {31}},\ \bibinfo
  {pages} {100755} (\bibinfo {year} {2021})},\ \Eprint
  {https://arxiv.org/abs/1906.08217} {arXiv:1906.08217 [astro-ph.CO]}
  \BibitemShut {NoStop}%
\bibitem [{\citenamefont {Gow}\ \emph {et~al.}(2020)\citenamefont {Gow},
  \citenamefont {Byrnes}, \citenamefont {Hall},\ and\ \citenamefont
  {Peacock}}]{Gow:2019pok}%
  \BibitemOpen
  \bibfield  {author} {\bibinfo {author} {\bibfnamefont {A.~D.}\ \bibnamefont
  {Gow}}, \bibinfo {author} {\bibfnamefont {C.~T.}\ \bibnamefont {Byrnes}},
  \bibinfo {author} {\bibfnamefont {A.}~\bibnamefont {Hall}},\ and\ \bibinfo
  {author} {\bibfnamefont {J.~A.}\ \bibnamefont {Peacock}},\ }\bibfield
  {title} {\bibinfo {title} {{Primordial black hole merger rates: distributions
  for multiple LIGO observables}},\ }\href
  {https://doi.org/10.1088/1475-7516/2020/01/031} {\bibfield  {journal}
  {\bibinfo  {journal} {JCAP}\ }\textbf {\bibinfo {volume} {01}},\ \bibinfo
  {pages} {031}},\ \Eprint {https://arxiv.org/abs/1911.12685} {arXiv:1911.12685
  [astro-ph.CO]} \BibitemShut {NoStop}%
\bibitem [{\citenamefont {Hall}\ \emph {et~al.}(2020)\citenamefont {Hall},
  \citenamefont {Gow},\ and\ \citenamefont {Byrnes}}]{Hall:2020daa}%
  \BibitemOpen
  \bibfield  {author} {\bibinfo {author} {\bibfnamefont {A.}~\bibnamefont
  {Hall}}, \bibinfo {author} {\bibfnamefont {A.~D.}\ \bibnamefont {Gow}},\ and\
  \bibinfo {author} {\bibfnamefont {C.~T.}\ \bibnamefont {Byrnes}},\ }\bibfield
   {title} {\bibinfo {title} {{Bayesian analysis of LIGO-Virgo mergers:
  Primordial vs. astrophysical black hole populations}},\ }\href
  {https://doi.org/10.1103/PhysRevD.102.123524} {\bibfield  {journal} {\bibinfo
   {journal} {Phys. Rev. D}\ }\textbf {\bibinfo {volume} {102}},\ \bibinfo
  {pages} {123524} (\bibinfo {year} {2020})},\ \Eprint
  {https://arxiv.org/abs/2008.13704} {arXiv:2008.13704 [astro-ph.CO]}
  \BibitemShut {NoStop}%
\bibitem [{\citenamefont {H\"utsi}\ \emph {et~al.}(2021)\citenamefont
  {H\"utsi}, \citenamefont {Raidal}, \citenamefont {Vaskonen},\ and\
  \citenamefont {Veerm\"ae}}]{Hutsi:2020sol}%
  \BibitemOpen
  \bibfield  {author} {\bibinfo {author} {\bibfnamefont {G.}~\bibnamefont
  {H\"utsi}}, \bibinfo {author} {\bibfnamefont {M.}~\bibnamefont {Raidal}},
  \bibinfo {author} {\bibfnamefont {V.}~\bibnamefont {Vaskonen}},\ and\
  \bibinfo {author} {\bibfnamefont {H.}~\bibnamefont {Veerm\"ae}},\ }\bibfield
  {title} {\bibinfo {title} {{Two populations of LIGO-Virgo black holes}},\
  }\href {https://doi.org/10.1088/1475-7516/2021/03/068} {\bibfield  {journal}
  {\bibinfo  {journal} {JCAP}\ }\textbf {\bibinfo {volume} {03}},\ \bibinfo
  {pages} {068}},\ \Eprint {https://arxiv.org/abs/2012.02786} {arXiv:2012.02786
  [astro-ph.CO]} \BibitemShut {NoStop}%
\bibitem [{\citenamefont {Clesse}\ and\ \citenamefont
  {Garcia-Bellido}(2022)}]{Clesse:2020ghq}%
  \BibitemOpen
  \bibfield  {author} {\bibinfo {author} {\bibfnamefont {S.}~\bibnamefont
  {Clesse}}\ and\ \bibinfo {author} {\bibfnamefont {J.}~\bibnamefont
  {Garcia-Bellido}},\ }\bibfield  {title} {\bibinfo {title} {{GW190425,
  GW190521 and GW190814: Three candidate mergers of primordial black holes from
  the QCD epoch}},\ }\href {https://doi.org/10.1016/j.dark.2022.101111}
  {\bibfield  {journal} {\bibinfo  {journal} {Phys. Dark Univ.}\ }\textbf
  {\bibinfo {volume} {38}},\ \bibinfo {pages} {101111} (\bibinfo {year}
  {2022})},\ \Eprint {https://arxiv.org/abs/2007.06481} {arXiv:2007.06481
  [astro-ph.CO]} \BibitemShut {NoStop}%
\bibitem [{\citenamefont {De~Luca}\ \emph
  {et~al.}(2020{\natexlab{a}})\citenamefont {De~Luca}, \citenamefont
  {Franciolini}, \citenamefont {Pani},\ and\ \citenamefont
  {Riotto}}]{DeLuca:2020qqa}%
  \BibitemOpen
  \bibfield  {author} {\bibinfo {author} {\bibfnamefont {V.}~\bibnamefont
  {De~Luca}}, \bibinfo {author} {\bibfnamefont {G.}~\bibnamefont
  {Franciolini}}, \bibinfo {author} {\bibfnamefont {P.}~\bibnamefont {Pani}},\
  and\ \bibinfo {author} {\bibfnamefont {A.}~\bibnamefont {Riotto}},\
  }\bibfield  {title} {\bibinfo {title} {{Primordial Black Holes Confront
  LIGO/Virgo data: Current situation}},\ }\href
  {https://doi.org/10.1088/1475-7516/2020/06/044} {\bibfield  {journal}
  {\bibinfo  {journal} {JCAP}\ }\textbf {\bibinfo {volume} {06}},\ \bibinfo
  {pages} {044}},\ \Eprint {https://arxiv.org/abs/2005.05641} {arXiv:2005.05641
  [astro-ph.CO]} \BibitemShut {NoStop}%
\bibitem [{\citenamefont {De~Luca}\ \emph {et~al.}(2021)\citenamefont
  {De~Luca}, \citenamefont {Desjacques}, \citenamefont {Franciolini},
  \citenamefont {Pani},\ and\ \citenamefont {Riotto}}]{DeLuca:2020sae}%
  \BibitemOpen
  \bibfield  {author} {\bibinfo {author} {\bibfnamefont {V.}~\bibnamefont
  {De~Luca}}, \bibinfo {author} {\bibfnamefont {V.}~\bibnamefont {Desjacques}},
  \bibinfo {author} {\bibfnamefont {G.}~\bibnamefont {Franciolini}}, \bibinfo
  {author} {\bibfnamefont {P.}~\bibnamefont {Pani}},\ and\ \bibinfo {author}
  {\bibfnamefont {A.}~\bibnamefont {Riotto}},\ }\bibfield  {title} {\bibinfo
  {title} {{GW190521 Mass Gap Event and the Primordial Black Hole Scenario}},\
  }\href {https://doi.org/10.1103/PhysRevLett.126.051101} {\bibfield  {journal}
  {\bibinfo  {journal} {Phys. Rev. Lett.}\ }\textbf {\bibinfo {volume} {126}},\
  \bibinfo {pages} {051101} (\bibinfo {year} {2021})},\ \Eprint
  {https://arxiv.org/abs/2009.01728} {arXiv:2009.01728 [astro-ph.CO]}
  \BibitemShut {NoStop}%
\bibitem [{\citenamefont {Jedamzik}(2021)}]{Jedamzik:2020omx}%
  \BibitemOpen
  \bibfield  {author} {\bibinfo {author} {\bibfnamefont {K.}~\bibnamefont
  {Jedamzik}},\ }\bibfield  {title} {\bibinfo {title} {{Consistency of
  Primordial Black Hole Dark Matter with LIGO/Virgo Merger Rates}},\ }\href
  {https://doi.org/10.1103/PhysRevLett.126.051302} {\bibfield  {journal}
  {\bibinfo  {journal} {Phys. Rev. Lett.}\ }\textbf {\bibinfo {volume} {126}},\
  \bibinfo {pages} {051302} (\bibinfo {year} {2021})},\ \Eprint
  {https://arxiv.org/abs/2007.03565} {arXiv:2007.03565 [astro-ph.CO]}
  \BibitemShut {NoStop}%
\bibitem [{\citenamefont {Jedamzik}(2020)}]{Jedamzik:2020ypm}%
  \BibitemOpen
  \bibfield  {author} {\bibinfo {author} {\bibfnamefont {K.}~\bibnamefont
  {Jedamzik}},\ }\bibfield  {title} {\bibinfo {title} {{Primordial Black Hole
  Dark Matter and the LIGO/Virgo observations}},\ }\href
  {https://doi.org/10.1088/1475-7516/2020/09/022} {\bibfield  {journal}
  {\bibinfo  {journal} {JCAP}\ }\textbf {\bibinfo {volume} {09}},\ \bibinfo
  {pages} {022}},\ \Eprint {https://arxiv.org/abs/2006.11172} {arXiv:2006.11172
  [astro-ph.CO]} \BibitemShut {NoStop}%
\bibitem [{\citenamefont {Wong}\ \emph {et~al.}(2021)\citenamefont {Wong},
  \citenamefont {Franciolini}, \citenamefont {De~Luca}, \citenamefont
  {Baibhav}, \citenamefont {Berti}, \citenamefont {Pani},\ and\ \citenamefont
  {Riotto}}]{Wong:2020yig}%
  \BibitemOpen
  \bibfield  {author} {\bibinfo {author} {\bibfnamefont {K.~W.~K.}\
  \bibnamefont {Wong}}, \bibinfo {author} {\bibfnamefont {G.}~\bibnamefont
  {Franciolini}}, \bibinfo {author} {\bibfnamefont {V.}~\bibnamefont
  {De~Luca}}, \bibinfo {author} {\bibfnamefont {V.}~\bibnamefont {Baibhav}},
  \bibinfo {author} {\bibfnamefont {E.}~\bibnamefont {Berti}}, \bibinfo
  {author} {\bibfnamefont {P.}~\bibnamefont {Pani}},\ and\ \bibinfo {author}
  {\bibfnamefont {A.}~\bibnamefont {Riotto}},\ }\bibfield  {title} {\bibinfo
  {title} {{Constraining the primordial black hole scenario with Bayesian
  inference and machine learning: the GWTC-2 gravitational wave catalog}},\
  }\href {https://doi.org/10.1103/PhysRevD.103.023026} {\bibfield  {journal}
  {\bibinfo  {journal} {Phys. Rev. D}\ }\textbf {\bibinfo {volume} {103}},\
  \bibinfo {pages} {023026} (\bibinfo {year} {2021})},\ \Eprint
  {https://arxiv.org/abs/2011.01865} {arXiv:2011.01865 [gr-qc]} \BibitemShut
  {NoStop}%
\bibitem [{\citenamefont {Escriv\`a}\ \emph {et~al.}(2022)\citenamefont
  {Escriv\`a}, \citenamefont {Bagui},\ and\ \citenamefont
  {Clesse}}]{Escriva:2022bwe}%
  \BibitemOpen
  \bibfield  {author} {\bibinfo {author} {\bibfnamefont {A.}~\bibnamefont
  {Escriv\`a}}, \bibinfo {author} {\bibfnamefont {E.}~\bibnamefont {Bagui}},\
  and\ \bibinfo {author} {\bibfnamefont {S.}~\bibnamefont {Clesse}},\
  }\bibfield  {title} {\bibinfo {title} {{Simulations of PBH formation at the
  QCD epoch and comparison with the GWTC-3 catalog}},\ }\href@noop {} {\
  (\bibinfo {year} {2022})},\ \Eprint {https://arxiv.org/abs/2209.06196}
  {arXiv:2209.06196 [astro-ph.CO]} \BibitemShut {NoStop}%
\bibitem [{\citenamefont {Franciolini}\ \emph {et~al.}(2022)\citenamefont
  {Franciolini}, \citenamefont {Musco}, \citenamefont {Pani},\ and\
  \citenamefont {Urbano}}]{Franciolini:2022tfm}%
  \BibitemOpen
  \bibfield  {author} {\bibinfo {author} {\bibfnamefont {G.}~\bibnamefont
  {Franciolini}}, \bibinfo {author} {\bibfnamefont {I.}~\bibnamefont {Musco}},
  \bibinfo {author} {\bibfnamefont {P.}~\bibnamefont {Pani}},\ and\ \bibinfo
  {author} {\bibfnamefont {A.}~\bibnamefont {Urbano}},\ }\bibfield  {title}
  {\bibinfo {title} {{From inflation to black hole mergers and back again:
  Gravitational-wave data-driven constraints on inflationary scenarios with a
  first-principle model of primordial black holes across the QCD epoch}},\
  }\href@noop {} {\  (\bibinfo {year} {2022})},\ \Eprint
  {https://arxiv.org/abs/2209.05959} {arXiv:2209.05959 [astro-ph.CO]}
  \BibitemShut {NoStop}%
\bibitem [{\citenamefont {{Ricotti}}(2007)}]{2007ApJ...662...53R}%
  \BibitemOpen
  \bibfield  {author} {\bibinfo {author} {\bibfnamefont {M.}~\bibnamefont
  {{Ricotti}}},\ }\bibfield  {title} {\bibinfo {title} {{Bondi Accretion in the
  Early Universe}},\ }\href {https://doi.org/10.1086/516562} {\bibfield
  {journal} {\bibinfo  {journal} {\apj}\ }\textbf {\bibinfo {volume} {662}},\
  \bibinfo {pages} {53} (\bibinfo {year} {2007})},\ \Eprint
  {https://arxiv.org/abs/0706.0864} {arXiv:0706.0864 [astro-ph]} \BibitemShut
  {NoStop}%
\bibitem [{\citenamefont {{Ricotti}}\ \emph {et~al.}(2008)\citenamefont
  {{Ricotti}}, \citenamefont {{Ostriker}},\ and\ \citenamefont
  {{Mack}}}]{2008ApJ...680..829R}%
  \BibitemOpen
  \bibfield  {author} {\bibinfo {author} {\bibfnamefont {M.}~\bibnamefont
  {{Ricotti}}}, \bibinfo {author} {\bibfnamefont {J.~P.}\ \bibnamefont
  {{Ostriker}}},\ and\ \bibinfo {author} {\bibfnamefont {K.~J.}\ \bibnamefont
  {{Mack}}},\ }\bibfield  {title} {\bibinfo {title} {{Effect of Primordial
  Black Holes on the Cosmic Microwave Background and Cosmological Parameter
  Estimates}},\ }\href {https://doi.org/10.1086/587831} {\bibfield  {journal}
  {\bibinfo  {journal} {\apj}\ }\textbf {\bibinfo {volume} {680}},\ \bibinfo
  {pages} {829} (\bibinfo {year} {2008})},\ \Eprint
  {https://arxiv.org/abs/0709.0524} {arXiv:0709.0524 [astro-ph]} \BibitemShut
  {NoStop}%
\bibitem [{\citenamefont {Ali-Haïmoud}\ and\ \citenamefont
  {Kamionkowski}(2017)}]{AliHaimoud2017Cosmic}%
  \BibitemOpen
  \bibfield  {author} {\bibinfo {author} {\bibfnamefont {Y.}~\bibnamefont
  {Ali-Haïmoud}}\ and\ \bibinfo {author} {\bibfnamefont {M.}~\bibnamefont
  {Kamionkowski}},\ }\bibfield  {title} {\bibinfo {title} {{Cosmic microwave
  background limits on accreting primordial black holes}},\ }\href
  {https://doi.org/10.1103/PhysRevD.95.043534} {\bibfield  {journal} {\bibinfo
  {journal} {Phys. Rev.}\ }\textbf {\bibinfo {volume} {D95}},\ \bibinfo {pages}
  {043534} (\bibinfo {year} {2017})},\ \Eprint
  {https://arxiv.org/abs/1612.05644} {arXiv:1612.05644 [astro-ph.CO]}
  \BibitemShut {NoStop}%
\bibitem [{\citenamefont {Poulin}\ \emph {et~al.}(2017)\citenamefont {Poulin},
  \citenamefont {Serpico}, \citenamefont {Calore}, \citenamefont {Clesse},\
  and\ \citenamefont {Kohri}}]{Poulin2017CMB}%
  \BibitemOpen
  \bibfield  {author} {\bibinfo {author} {\bibfnamefont {V.}~\bibnamefont
  {Poulin}}, \bibinfo {author} {\bibfnamefont {P.~D.}\ \bibnamefont {Serpico}},
  \bibinfo {author} {\bibfnamefont {F.}~\bibnamefont {Calore}}, \bibinfo
  {author} {\bibfnamefont {S.}~\bibnamefont {Clesse}},\ and\ \bibinfo {author}
  {\bibfnamefont {K.}~\bibnamefont {Kohri}},\ }\bibfield  {title} {\bibinfo
  {title} {{CMB bounds on disk-accreting massive primordial black holes}},\
  }\href {https://doi.org/10.1103/PhysRevD.96.083524} {\bibfield  {journal}
  {\bibinfo  {journal} {Phys. Rev.}\ }\textbf {\bibinfo {volume} {D96}},\
  \bibinfo {pages} {083524} (\bibinfo {year} {2017})},\ \Eprint
  {https://arxiv.org/abs/1707.04206} {arXiv:1707.04206 [astro-ph.CO]}
  \BibitemShut {NoStop}%
\bibitem [{\citenamefont {Aloni}\ \emph {et~al.}(2017)\citenamefont {Aloni},
  \citenamefont {Blum},\ and\ \citenamefont {Flauger}}]{Aloni:2016kuh}%
  \BibitemOpen
  \bibfield  {author} {\bibinfo {author} {\bibfnamefont {D.}~\bibnamefont
  {Aloni}}, \bibinfo {author} {\bibfnamefont {K.}~\bibnamefont {Blum}},\ and\
  \bibinfo {author} {\bibfnamefont {R.}~\bibnamefont {Flauger}},\ }\bibfield
  {title} {\bibinfo {title} {{Cosmic microwave background constraints on
  primordial black hole dark matter}},\ }\href
  {https://doi.org/10.1088/1475-7516/2017/05/017} {\bibfield  {journal}
  {\bibinfo  {journal} {JCAP}\ }\textbf {\bibinfo {volume} {05}},\ \bibinfo
  {pages} {017}},\ \Eprint {https://arxiv.org/abs/1612.06811} {arXiv:1612.06811
  [astro-ph.CO]} \BibitemShut {NoStop}%
\bibitem [{\citenamefont {Serpico}\ \emph {et~al.}(2020)\citenamefont
  {Serpico}, \citenamefont {Poulin}, \citenamefont {Inman},\ and\ \citenamefont
  {Kohri}}]{Serpico2020Cosmic}%
  \BibitemOpen
  \bibfield  {author} {\bibinfo {author} {\bibfnamefont {P.~D.}\ \bibnamefont
  {Serpico}}, \bibinfo {author} {\bibfnamefont {V.}~\bibnamefont {Poulin}},
  \bibinfo {author} {\bibfnamefont {D.}~\bibnamefont {Inman}},\ and\ \bibinfo
  {author} {\bibfnamefont {K.}~\bibnamefont {Kohri}},\ }\bibfield  {title}
  {\bibinfo {title} {{Cosmic microwave background bounds on primordial black
  holes including dark matter halo accretion}},\ }\href
  {https://doi.org/10.1103/PhysRevResearch.2.023204} {\bibfield  {journal}
  {\bibinfo  {journal} {Phys. Rev. Res.}\ }\textbf {\bibinfo {volume} {2}},\
  \bibinfo {pages} {023204} (\bibinfo {year} {2020})},\ \Eprint
  {https://arxiv.org/abs/2002.10771} {arXiv:2002.10771 [astro-ph.CO]}
  \BibitemShut {NoStop}%
\bibitem [{\citenamefont {Juan}\ \emph {et~al.}(2022)\citenamefont {Juan},
  \citenamefont {Serpico},\ and\ \citenamefont
  {Franco~Abell\'an}}]{Juan:2022mir}%
  \BibitemOpen
  \bibfield  {author} {\bibinfo {author} {\bibfnamefont {J.~I.}\ \bibnamefont
  {Juan}}, \bibinfo {author} {\bibfnamefont {P.~D.}\ \bibnamefont {Serpico}},\
  and\ \bibinfo {author} {\bibfnamefont {G.}~\bibnamefont {Franco~Abell\'an}},\
  }\bibfield  {title} {\bibinfo {title} {{The QCD phase transition behind a PBH
  origin of LIGO/Virgo events?}},\ }\href
  {https://doi.org/10.1088/1475-7516/2022/07/009} {\bibfield  {journal}
  {\bibinfo  {journal} {JCAP}\ }\textbf {\bibinfo {volume} {07}}\bibfield
  {number} {\bibinfo  {number} { (07)},\ \bibinfo {pages} {009}},\ }\Eprint
  {https://arxiv.org/abs/2204.07027} {arXiv:2204.07027 [astro-ph.CO]}
  \BibitemShut {NoStop}%
\bibitem [{\citenamefont {Piga}\ \emph {et~al.}(2022)\citenamefont {Piga},
  \citenamefont {Lucca}, \citenamefont {Bellomo}, \citenamefont {Bosch-Ramon},
  \citenamefont {Matarrese}, \citenamefont {Raccanelli},\ and\ \citenamefont
  {Verde}}]{Piga:2022ysp}%
  \BibitemOpen
  \bibfield  {author} {\bibinfo {author} {\bibfnamefont {L.}~\bibnamefont
  {Piga}}, \bibinfo {author} {\bibfnamefont {M.}~\bibnamefont {Lucca}},
  \bibinfo {author} {\bibfnamefont {N.}~\bibnamefont {Bellomo}}, \bibinfo
  {author} {\bibfnamefont {V.}~\bibnamefont {Bosch-Ramon}}, \bibinfo {author}
  {\bibfnamefont {S.}~\bibnamefont {Matarrese}}, \bibinfo {author}
  {\bibfnamefont {A.}~\bibnamefont {Raccanelli}},\ and\ \bibinfo {author}
  {\bibfnamefont {L.}~\bibnamefont {Verde}},\ }\bibfield  {title} {\bibinfo
  {title} {{The effect of outflows on CMB bounds from Primordial Black Hole
  accretion}},\ }\href@noop {} {\  (\bibinfo {year} {2022})},\ \Eprint
  {https://arxiv.org/abs/2210.14934} {arXiv:2210.14934 [astro-ph.CO]}
  \BibitemShut {NoStop}%
\bibitem [{\citenamefont {{Mack}}\ \emph {et~al.}(2007)\citenamefont {{Mack}},
  \citenamefont {{Ostriker}},\ and\ \citenamefont
  {{Ricotti}}}]{2007ApJ...665.1277M}%
  \BibitemOpen
  \bibfield  {author} {\bibinfo {author} {\bibfnamefont {K.~J.}\ \bibnamefont
  {{Mack}}}, \bibinfo {author} {\bibfnamefont {J.~P.}\ \bibnamefont
  {{Ostriker}}},\ and\ \bibinfo {author} {\bibfnamefont {M.}~\bibnamefont
  {{Ricotti}}},\ }\bibfield  {title} {\bibinfo {title} {{Growth of Structure
  Seeded by Primordial Black Holes}},\ }\href {https://doi.org/10.1086/518998}
  {\bibfield  {journal} {\bibinfo  {journal} {\apj}\ }\textbf {\bibinfo
  {volume} {665}},\ \bibinfo {pages} {1277} (\bibinfo {year} {2007})},\ \Eprint
  {https://arxiv.org/abs/astro-ph/0608642} {arXiv:astro-ph/0608642 [astro-ph]}
  \BibitemShut {NoStop}%
\bibitem [{\citenamefont {Park}\ and\ \citenamefont
  {Ricotti}(2013)}]{Park:2012cr}%
  \BibitemOpen
  \bibfield  {author} {\bibinfo {author} {\bibfnamefont {K.}~\bibnamefont
  {Park}}\ and\ \bibinfo {author} {\bibfnamefont {M.}~\bibnamefont {Ricotti}},\
  }\bibfield  {title} {\bibinfo {title} {{Accretion onto Black Holes from Large
  Scales Regulated by Radiative Feedback. III. Enhanced Luminosity of
  Intermediate Mass Black Holes Moving at Supersonic Speeds}},\ }\href
  {https://doi.org/10.1088/0004-637X/767/2/163} {\bibfield  {journal} {\bibinfo
   {journal} {Astrophys. J.}\ }\textbf {\bibinfo {volume} {767}},\ \bibinfo
  {pages} {163} (\bibinfo {year} {2013})},\ \Eprint
  {https://arxiv.org/abs/1211.0542} {arXiv:1211.0542 [astro-ph.CO]}
  \BibitemShut {NoStop}%
\bibitem [{\citenamefont {Sugimura}\ and\ \citenamefont
  {Ricotti}(2020)}]{Sugimura:2020rdw}%
  \BibitemOpen
  \bibfield  {author} {\bibinfo {author} {\bibfnamefont {K.}~\bibnamefont
  {Sugimura}}\ and\ \bibinfo {author} {\bibfnamefont {M.}~\bibnamefont
  {Ricotti}},\ }\bibfield  {title} {\bibinfo {title} {{Structure and
  Instability of the Ionization Fronts around Moving Black Holes}},\ }\href
  {https://doi.org/10.1093/mnras/staa1394} {\bibfield  {journal} {\bibinfo
  {journal} {Mon. Not. Roy. Astron. Soc.}\ }\textbf {\bibinfo {volume} {495}},\
  \bibinfo {pages} {2966} (\bibinfo {year} {2020})},\ \Eprint
  {https://arxiv.org/abs/2003.05625} {arXiv:2003.05625 [astro-ph.GA]}
  \BibitemShut {NoStop}%
\bibitem [{\citenamefont {Inman}\ and\ \citenamefont
  {Ali-Ha\"\i{}moud}(2019)}]{Inman2019Early}%
  \BibitemOpen
  \bibfield  {author} {\bibinfo {author} {\bibfnamefont {D.}~\bibnamefont
  {Inman}}\ and\ \bibinfo {author} {\bibfnamefont {Y.}~\bibnamefont
  {Ali-Ha\"\i{}moud}},\ }\bibfield  {title} {\bibinfo {title} {{Early structure
  formation in primordial black hole cosmologies}},\ }\href
  {https://doi.org/10.1103/PhysRevD.100.083528} {\bibfield  {journal} {\bibinfo
   {journal} {Phys. Rev. D}\ }\textbf {\bibinfo {volume} {100}},\ \bibinfo
  {pages} {083528} (\bibinfo {year} {2019})},\ \Eprint
  {https://arxiv.org/abs/1907.08129} {arXiv:1907.08129 [astro-ph.CO]}
  \BibitemShut {NoStop}%
\bibitem [{\citenamefont {Scarcella}\ \emph {et~al.}(2021)\citenamefont
  {Scarcella}, \citenamefont {Gaggero}, \citenamefont {Connors}, \citenamefont
  {Manshanden}, \citenamefont {Ricotti},\ and\ \citenamefont
  {Bertone}}]{Scarcella:2020ssk}%
  \BibitemOpen
  \bibfield  {author} {\bibinfo {author} {\bibfnamefont {F.}~\bibnamefont
  {Scarcella}}, \bibinfo {author} {\bibfnamefont {D.}~\bibnamefont {Gaggero}},
  \bibinfo {author} {\bibfnamefont {R.}~\bibnamefont {Connors}}, \bibinfo
  {author} {\bibfnamefont {J.}~\bibnamefont {Manshanden}}, \bibinfo {author}
  {\bibfnamefont {M.}~\bibnamefont {Ricotti}},\ and\ \bibinfo {author}
  {\bibfnamefont {G.}~\bibnamefont {Bertone}},\ }\bibfield  {title} {\bibinfo
  {title} {{Multiwavelength detectability of isolated black holes in the Milky
  Way}},\ }\href {https://doi.org/10.1093/mnras/stab1533} {\bibfield  {journal}
  {\bibinfo  {journal} {Mon. Not. Roy. Astron. Soc.}\ }\textbf {\bibinfo
  {volume} {505}},\ \bibinfo {pages} {4036} (\bibinfo {year} {2021})},\ \Eprint
  {https://arxiv.org/abs/2012.10421} {arXiv:2012.10421 [astro-ph.HE]}
  \BibitemShut {NoStop}%
\bibitem [{\citenamefont {Slatyer}\ and\ \citenamefont
  {Wu}(2017)}]{Slatyer2016General}%
  \BibitemOpen
  \bibfield  {author} {\bibinfo {author} {\bibfnamefont {T.~R.}\ \bibnamefont
  {Slatyer}}\ and\ \bibinfo {author} {\bibfnamefont {C.-L.}\ \bibnamefont
  {Wu}},\ }\bibfield  {title} {\bibinfo {title} {{General Constraints on Dark
  Matter Decay from the Cosmic Microwave Background}},\ }\href
  {https://doi.org/10.1103/PhysRevD.95.023010} {\bibfield  {journal} {\bibinfo
  {journal} {Phys. Rev. D}\ }\textbf {\bibinfo {volume} {95}},\ \bibinfo
  {pages} {023010} (\bibinfo {year} {2017})},\ \Eprint
  {https://arxiv.org/abs/1610.06933} {arXiv:1610.06933 [astro-ph.CO]}
  \BibitemShut {NoStop}%
\bibitem [{\citenamefont {Slatyer}(2016)}]{Slatyer2015IndirectI}%
  \BibitemOpen
  \bibfield  {author} {\bibinfo {author} {\bibfnamefont {T.~R.}\ \bibnamefont
  {Slatyer}},\ }\bibfield  {title} {\bibinfo {title} {{Indirect dark matter
  signatures in the cosmic dark ages. I. Generalizing the bound on s-wave dark
  matter annihilation from Planck results}},\ }\href
  {https://doi.org/10.1103/PhysRevD.93.023527} {\bibfield  {journal} {\bibinfo
  {journal} {Phys. Rev. D}\ }\textbf {\bibinfo {volume} {93}},\ \bibinfo
  {pages} {023527} (\bibinfo {year} {2016})},\ \Eprint
  {https://arxiv.org/abs/1506.03811} {arXiv:1506.03811 [hep-ph]} \BibitemShut
  {NoStop}%
\bibitem [{\citenamefont {Lucca}\ \emph {et~al.}(2020)\citenamefont {Lucca},
  \citenamefont {Sch\"oneberg}, \citenamefont {Hooper}, \citenamefont
  {Lesgourgues},\ and\ \citenamefont {Chluba}}]{Lucca2019Synergy}%
  \BibitemOpen
  \bibfield  {author} {\bibinfo {author} {\bibfnamefont {M.}~\bibnamefont
  {Lucca}}, \bibinfo {author} {\bibfnamefont {N.}~\bibnamefont {Sch\"oneberg}},
  \bibinfo {author} {\bibfnamefont {D.~C.}\ \bibnamefont {Hooper}}, \bibinfo
  {author} {\bibfnamefont {J.}~\bibnamefont {Lesgourgues}},\ and\ \bibinfo
  {author} {\bibfnamefont {J.}~\bibnamefont {Chluba}},\ }\bibfield  {title}
  {\bibinfo {title} {{The synergy between CMB spectral distortions and
  anisotropies}},\ }\href {https://doi.org/10.1088/1475-7516/2020/02/026}
  {\bibfield  {journal} {\bibinfo  {journal} {JCAP}\ }\textbf {\bibinfo
  {volume} {02}},\ \bibinfo {pages} {026}},\ \Eprint
  {https://arxiv.org/abs/1910.04619} {arXiv:1910.04619 [astro-ph.CO]}
  \BibitemShut {NoStop}%
\bibitem [{\citenamefont {Ali-Haimoud}\ and\ \citenamefont
  {Hirata}(2011)}]{AliHaimoud2010HyRec}%
  \BibitemOpen
  \bibfield  {author} {\bibinfo {author} {\bibfnamefont {Y.}~\bibnamefont
  {Ali-Haimoud}}\ and\ \bibinfo {author} {\bibfnamefont {C.~M.}\ \bibnamefont
  {Hirata}},\ }\bibfield  {title} {\bibinfo {title} {{HyRec: A fast and highly
  accurate primordial hydrogen and helium recombination code}},\ }\href
  {https://doi.org/10.1103/PhysRevD.83.043513} {\bibfield  {journal} {\bibinfo
  {journal} {Phys. Rev.}\ }\textbf {\bibinfo {volume} {D83}},\ \bibinfo {pages}
  {043513} (\bibinfo {year} {2011})},\ \Eprint
  {https://arxiv.org/abs/1011.3758} {arXiv:1011.3758 [astro-ph.CO]}
  \BibitemShut {NoStop}%
\bibitem [{\citenamefont {Lee}\ and\ \citenamefont
  {Ali-Ha\"\i{}moud}(2020)}]{Lee2020HYREC}%
  \BibitemOpen
  \bibfield  {author} {\bibinfo {author} {\bibfnamefont {N.}~\bibnamefont
  {Lee}}\ and\ \bibinfo {author} {\bibfnamefont {Y.}~\bibnamefont
  {Ali-Ha\"\i{}moud}},\ }\bibfield  {title} {\bibinfo {title} {{HYREC-2: a
  highly accurate sub-millisecond recombination code}},\ }\href
  {https://doi.org/10.1103/PhysRevD.102.083517} {\bibfield  {journal} {\bibinfo
   {journal} {Phys. Rev. D}\ }\textbf {\bibinfo {volume} {102}},\ \bibinfo
  {pages} {083517} (\bibinfo {year} {2020})},\ \Eprint
  {https://arxiv.org/abs/2007.14114} {arXiv:2007.14114 [astro-ph.CO]}
  \BibitemShut {NoStop}%
\bibitem [{\citenamefont {Galli}\ \emph {et~al.}(2013)\citenamefont {Galli},
  \citenamefont {Slatyer}, \citenamefont {Valdes},\ and\ \citenamefont
  {Iocco}}]{Galli2013Systematic}%
  \BibitemOpen
  \bibfield  {author} {\bibinfo {author} {\bibfnamefont {S.}~\bibnamefont
  {Galli}}, \bibinfo {author} {\bibfnamefont {T.~R.}\ \bibnamefont {Slatyer}},
  \bibinfo {author} {\bibfnamefont {M.}~\bibnamefont {Valdes}},\ and\ \bibinfo
  {author} {\bibfnamefont {F.}~\bibnamefont {Iocco}},\ }\bibfield  {title}
  {\bibinfo {title} {{Systematic Uncertainties In Constraining Dark Matter
  Annihilation From The Cosmic Microwave Background}},\ }\href
  {https://doi.org/10.1103/PhysRevD.88.063502} {\bibfield  {journal} {\bibinfo
  {journal} {Phys. Rev.}\ }\textbf {\bibinfo {volume} {D88}},\ \bibinfo {pages}
  {063502} (\bibinfo {year} {2013})},\ \Eprint
  {https://arxiv.org/abs/1306.0563} {arXiv:1306.0563 [astro-ph.CO]}
  \BibitemShut {NoStop}%
\bibitem [{\citenamefont {Lesgourgues}(2011)}]{Lesgourgues2011CosmicI}%
  \BibitemOpen
  \bibfield  {author} {\bibinfo {author} {\bibfnamefont {J.}~\bibnamefont
  {Lesgourgues}},\ }\bibfield  {title} {\bibinfo {title} {{The Cosmic Linear
  Anisotropy Solving System (CLASS) I: Overview}},\ }\href@noop {} {\
  (\bibinfo {year} {2011})},\ \Eprint {https://arxiv.org/abs/1104.2932}
  {arXiv:1104.2932 [astro-ph.IM]} \BibitemShut {NoStop}%
\bibitem [{\citenamefont {Blas}\ \emph {et~al.}(2011)\citenamefont {Blas},
  \citenamefont {Lesgourgues},\ and\ \citenamefont {Tram}}]{Blas2011Cosmic}%
  \BibitemOpen
  \bibfield  {author} {\bibinfo {author} {\bibfnamefont {D.}~\bibnamefont
  {Blas}}, \bibinfo {author} {\bibfnamefont {J.}~\bibnamefont {Lesgourgues}},\
  and\ \bibinfo {author} {\bibfnamefont {T.}~\bibnamefont {Tram}},\ }\bibfield
  {title} {\bibinfo {title} {{The Cosmic Linear Anisotropy Solving System
  (CLASS) II: Approximation schemes}},\ }\href
  {https://doi.org/10.1088/1475-7516/2011/07/034} {\bibfield  {journal}
  {\bibinfo  {journal} {JCAP}\ }\textbf {\bibinfo {volume} {1107}},\ \bibinfo
  {pages} {034}},\ \Eprint {https://arxiv.org/abs/1104.2933} {arXiv:1104.2933
  [astro-ph.CO]} \BibitemShut {NoStop}%
\bibitem [{\citenamefont {{St\"ocker, P. and Kr\"amer, M. and Lesgourgues, J.
  and Poulin, V.}}(2018)}]{Stocker2018Exotic}%
  \BibitemOpen
  \bibfield  {author} {\bibinfo {author} {\bibnamefont {{St\"ocker, P. and
  Kr\"amer, M. and Lesgourgues, J. and Poulin, V.}}},\ }\bibfield  {title}
  {\bibinfo {title} {{Exotic energy injection with ExoCLASS: Application to the
  Higgs portal model and evaporating black holes}},\ }\href
  {https://doi.org/10.1088/1475-7516/2018/03/018} {\bibfield  {journal}
  {\bibinfo  {journal} {JCAP}\ }\textbf {\bibinfo {volume} {1803}}\bibfield
  {number} {\bibinfo  {number} { (03)},\ \bibinfo {pages} {018}},\ }\Eprint
  {https://arxiv.org/abs/1801.01871} {arXiv:1801.01871 [astro-ph.CO]}
  \BibitemShut {NoStop}%
\bibitem [{\citenamefont {Audren}\ \emph {et~al.}(2013)\citenamefont {Audren},
  \citenamefont {Lesgourgues}, \citenamefont {Benabed},\ and\ \citenamefont
  {Prunet}}]{Audren2013Conservative}%
  \BibitemOpen
  \bibfield  {author} {\bibinfo {author} {\bibfnamefont {B.}~\bibnamefont
  {Audren}}, \bibinfo {author} {\bibfnamefont {J.}~\bibnamefont {Lesgourgues}},
  \bibinfo {author} {\bibfnamefont {K.}~\bibnamefont {Benabed}},\ and\ \bibinfo
  {author} {\bibfnamefont {S.}~\bibnamefont {Prunet}},\ }\bibfield  {title}
  {\bibinfo {title} {{Conservative Constraints on Early Cosmology: an
  illustration of the Monte Python cosmological parameter inference code}},\
  }\href {https://doi.org/10.1088/1475-7516/2013/02/001} {\bibfield  {journal}
  {\bibinfo  {journal} {JCAP}\ }\textbf {\bibinfo {volume} {1302}},\ \bibinfo
  {pages} {001}},\ \Eprint {https://arxiv.org/abs/1210.7183} {arXiv:1210.7183
  [astro-ph.CO]} \BibitemShut {NoStop}%
\bibitem [{\citenamefont {Brinckmann}\ and\ \citenamefont
  {Lesgourgues}(2018)}]{Brinckmann2018MontePython}%
  \BibitemOpen
  \bibfield  {author} {\bibinfo {author} {\bibfnamefont {T.}~\bibnamefont
  {Brinckmann}}\ and\ \bibinfo {author} {\bibfnamefont {J.}~\bibnamefont
  {Lesgourgues}},\ }\bibfield  {title} {\bibinfo {title} {{MontePython 3:
  boosted MCMC sampler and other features}},\ }\href@noop {} {\  (\bibinfo
  {year} {2018})},\ \Eprint {https://arxiv.org/abs/1804.07261}
  {arXiv:1804.07261 [astro-ph.CO]} \BibitemShut {NoStop}%
\bibitem [{\citenamefont {Aghanim}\ \emph
  {et~al.}(2020{\natexlab{a}})\citenamefont {Aghanim} \emph
  {et~al.}}]{Aghanim2018PlanckVI}%
  \BibitemOpen
  \bibfield  {author} {\bibinfo {author} {\bibfnamefont {N.}~\bibnamefont
  {Aghanim}} \emph {et~al.} (\bibinfo {collaboration} {Planck}),\ }\bibfield
  {title} {\bibinfo {title} {{Planck 2018 results. VI. Cosmological
  parameters}},\ }\href {https://doi.org/10.1051/0004-6361/201833910}
  {\bibfield  {journal} {\bibinfo  {journal} {Astron. Astrophys.}\ }\textbf
  {\bibinfo {volume} {641}},\ \bibinfo {pages} {A6} (\bibinfo {year}
  {2020}{\natexlab{a}})},\ \Eprint {https://arxiv.org/abs/1807.06209}
  {arXiv:1807.06209 [astro-ph.CO]} \BibitemShut {NoStop}%
\bibitem [{\citenamefont {Aghanim}\ \emph
  {et~al.}(2020{\natexlab{b}})\citenamefont {Aghanim} \emph
  {et~al.}}]{Aghanim2018PlanckV}%
  \BibitemOpen
  \bibfield  {author} {\bibinfo {author} {\bibfnamefont {N.}~\bibnamefont
  {Aghanim}} \emph {et~al.} (\bibinfo {collaboration} {Planck}),\ }\bibfield
  {title} {\bibinfo {title} {{Planck 2018 results. V. CMB power spectra and
  likelihoods}},\ }\href {https://doi.org/10.1051/0004-6361/201936386}
  {\bibfield  {journal} {\bibinfo  {journal} {Astron. Astrophys.}\ }\textbf
  {\bibinfo {volume} {641}},\ \bibinfo {pages} {A5} (\bibinfo {year}
  {2020}{\natexlab{b}})},\ \Eprint {https://arxiv.org/abs/1907.12875}
  {arXiv:1907.12875 [astro-ph.CO]} \BibitemShut {NoStop}%
\bibitem [{\citenamefont {Gelman}\ and\ \citenamefont
  {Rubin}(1992)}]{Gelman1992Inference}%
  \BibitemOpen
  \bibfield  {author} {\bibinfo {author} {\bibfnamefont {A.}~\bibnamefont
  {Gelman}}\ and\ \bibinfo {author} {\bibfnamefont {D.~B.}\ \bibnamefont
  {Rubin}},\ }\bibfield  {title} {\bibinfo {title} {Inference from iterative
  simulation using multiple sequences},\ }\href
  {https://doi.org/10.1214/ss/1177011136} {\bibfield  {journal} {\bibinfo
  {journal} {Statist. Sci.}\ }\textbf {\bibinfo {volume} {7}},\ \bibinfo
  {pages} {457} (\bibinfo {year} {1992})}\BibitemShut {NoStop}%
\bibitem [{\citenamefont {Brandt}(2016)}]{Brandt:2016aco}%
  \BibitemOpen
  \bibfield  {author} {\bibinfo {author} {\bibfnamefont {T.~D.}\ \bibnamefont
  {Brandt}},\ }\bibfield  {title} {\bibinfo {title} {{Constraints on MACHO Dark
  Matter from Compact Stellar Systems in Ultra-Faint Dwarf Galaxies}},\ }\href
  {https://doi.org/10.3847/2041-8205/824/2/L31} {\bibfield  {journal} {\bibinfo
   {journal} {Astrophys. J. Lett.}\ }\textbf {\bibinfo {volume} {824}},\
  \bibinfo {pages} {L31} (\bibinfo {year} {2016})},\ \Eprint
  {https://arxiv.org/abs/1605.03665} {arXiv:1605.03665 [astro-ph.GA]}
  \BibitemShut {NoStop}%
\bibitem [{\citenamefont {Ziparo}\ \emph {et~al.}(2022)\citenamefont {Ziparo},
  \citenamefont {Gallerani}, \citenamefont {Ferrara},\ and\ \citenamefont
  {Vito}}]{Ziparo:2022fnc}%
  \BibitemOpen
  \bibfield  {author} {\bibinfo {author} {\bibfnamefont {F.}~\bibnamefont
  {Ziparo}}, \bibinfo {author} {\bibfnamefont {S.}~\bibnamefont {Gallerani}},
  \bibinfo {author} {\bibfnamefont {A.}~\bibnamefont {Ferrara}},\ and\ \bibinfo
  {author} {\bibfnamefont {F.}~\bibnamefont {Vito}},\ }\bibfield  {title}
  {\bibinfo {title} {{Cosmic radiation backgrounds from primordial black
  holes}},\ }\href {https://doi.org/10.1093/mnras/stac2705} {\bibfield
  {journal} {\bibinfo  {journal} {Mon. Not. Roy. Astron. Soc.}\ }\textbf
  {\bibinfo {volume} {517}},\ \bibinfo {pages} {1086} (\bibinfo {year}
  {2022})},\ \Eprint {https://arxiv.org/abs/2209.09907} {arXiv:2209.09907
  [astro-ph.CO]} \BibitemShut {NoStop}%
\bibitem [{\citenamefont {Mena}\ \emph {et~al.}(2019)\citenamefont {Mena},
  \citenamefont {Palomares-Ruiz}, \citenamefont {Villanueva-Domingo},\ and\
  \citenamefont {Witte}}]{Mena:2019nhm}%
  \BibitemOpen
  \bibfield  {author} {\bibinfo {author} {\bibfnamefont {O.}~\bibnamefont
  {Mena}}, \bibinfo {author} {\bibfnamefont {S.}~\bibnamefont
  {Palomares-Ruiz}}, \bibinfo {author} {\bibfnamefont {P.}~\bibnamefont
  {Villanueva-Domingo}},\ and\ \bibinfo {author} {\bibfnamefont {S.~J.}\
  \bibnamefont {Witte}},\ }\bibfield  {title} {\bibinfo {title} {{Constraining
  the primordial black hole abundance with 21-cm cosmology}},\ }\href
  {https://doi.org/10.1103/PhysRevD.100.043540} {\bibfield  {journal} {\bibinfo
   {journal} {Phys. Rev. D}\ }\textbf {\bibinfo {volume} {100}},\ \bibinfo
  {pages} {043540} (\bibinfo {year} {2019})},\ \Eprint
  {https://arxiv.org/abs/1906.07735} {arXiv:1906.07735 [astro-ph.CO]}
  \BibitemShut {NoStop}%
\bibitem [{\citenamefont {De~Luca}\ \emph
  {et~al.}(2020{\natexlab{b}})\citenamefont {De~Luca}, \citenamefont
  {Franciolini}, \citenamefont {Pani},\ and\ \citenamefont
  {Riotto}}]{DeLuca2020Constraints}%
  \BibitemOpen
  \bibfield  {author} {\bibinfo {author} {\bibfnamefont {V.}~\bibnamefont
  {De~Luca}}, \bibinfo {author} {\bibfnamefont {G.}~\bibnamefont
  {Franciolini}}, \bibinfo {author} {\bibfnamefont {P.}~\bibnamefont {Pani}},\
  and\ \bibinfo {author} {\bibfnamefont {A.}~\bibnamefont {Riotto}},\
  }\bibfield  {title} {\bibinfo {title} {{Constraints on Primordial Black
  Holes: the Importance of Accretion}},\ }\href
  {https://doi.org/10.1103/PhysRevD.102.043505} {\bibfield  {journal} {\bibinfo
   {journal} {Phys. Rev. D}\ }\textbf {\bibinfo {volume} {102}},\ \bibinfo
  {pages} {043505} (\bibinfo {year} {2020}{\natexlab{b}})},\ \Eprint
  {https://arxiv.org/abs/2003.12589} {arXiv:2003.12589 [astro-ph.CO]}
  \BibitemShut {NoStop}%
\end{thebibliography}%

\end{document}